\documentclass[fleqn,usenatbib]{mnras}
\usepackage{makecell}

\setcellgapes{4pt}
\usepackage{gensymb}

\usepackage[T1]{fontenc}
\usepackage[dvipsnames]{xcolor}
\usepackage{soul}
\usepackage{tabularx}
\DeclareRobustCommand{\VAN}[3]{#2}
\let\VANthebibliography\thebibliography
\def\thebibliography{\DeclareRobustCommand{\VAN}[3]{##3}\VANthebibliography}

\newlength\mylength
\usepackage{graphicx}	
\usepackage{amsmath}	
\usepackage{amssymb}	
\usepackage{subfigure}
\newcommand{\RNum}[1]{\uppercase\expandafter{\romannumeral #1\relax}}
\usepackage{newtxtext,newtxmath}

\title{The MeerKAT Galaxy Clusters Legacy Survey: star formation
in massive clusters at $0.15 < z < 0.35$}

\author[K. C. Kesebonye]{K. C. Kesebonye$^{1,2}$
\thanks{Contact e-mail: \href{mailto:kabelokes.astro@gmail.com}{kabelokes.astro@gmail.com}}%
, M. Hilton$^{1,2,3}$,
K. Knowles$^{1,4,5}$,
W. D. Cotton$^{5,6}$,
T. E. Clarke$^{7}$,
S. I. Loubser$^{8}$, \newauthor
K. Moodley$^{1,2}$,
S. P. Sikhosana$^{1,2}$
\\
$^{1}$Astrophysics Research Centre, University of KwaZulu-Natal, Durban 4041, South Africa\\
$^{2}$School of Mathematics, Statistics, and Computer Science, 
University of KwaZulu-Natal, Westville 3696, South Africa\\
$^{3}$Wits Centre for Astrophysics, School of Physics, University of the Witwatersrand, Private Bag 3, 2050, Johannesburg, South Africa\\
$^{4}$Centre for Radio Astronomy Techniques and Technologies, Department of Physics and Electronics, 
Rhodes University, P.O. Box 94, Makhanda 6140, South Africa\\
$^{5}$South African Radio Astronomy Observatory, 2 Fir Street, Observatory 7925, South Africa\\
$^{6}$National Radio Astronomy Observatory, Charlottesville, VA 22903, USA\\
$^{7}$Naval Research Laboratory, Washington, DC 20375, USA\\
$^{8}$Centre for Space Research, North-West University, Potchefstroom, 2520, South Africa}

\date{Accepted XXX. Received YYY; in original form ZZZ}

\pubyear{2022}

\begin{document}
\label{firstpage}
\pagerange{\pageref{firstpage}--\pageref{lastpage}}
\maketitle

\begin{abstract}
We investigate dust-unbiased star formation rates 
(SFR) as a function of the environment in 20 massive clusters
($M_{200}>4\times10^{14}\,{\rm M}_{\odot}$)
between $0.15<z<0.35$ using radio luminosities ($L_{\rm 1.4GHz}$)
from the recently released MeerKAT Galaxy 
Cluster Legacy Survey catalogue. We use optical 
data from the Dark Energy Camera Legacy Survey to estimate
photo-$z$s and assign cluster membership. We observe a steady decline
in the fraction ($f_{\rm SF}$) of star-forming galaxies from $2R_{200}$ to the 
cluster centres in our full cluster sample, 
but notice a significant difference in $f_{\rm SF}$
gradients between clusters hosting large-scale extended radio 
emission in the form of haloes and relics
(associated with ongoing merger activity) and non-radio-halo/relic hosting clusters. 
For star-forming galaxies within $R_{200}$, the $f_{\rm SF}$ in 
clusters hosting radio haloes and relics ($0.148\pm0.016$) 
is $\approx23\%$ higher than in non-radio-halo/relic hosting clusters ($0.120\pm0.011$).
We observe a $3\sigma$ difference between the total SFR 
normalised by cluster mass for non-radio-halo/relic hosting clusters ($21.5\pm1.9$\,M$_{\odot}$yr$^{-1}$/$10^{14}$M$_{\odot}$) 
and for clusters with radio haloes and relics 
($26.1\pm1.4$\,M$_{\odot}$yr$^{-1}$/$10^{14}$M$_{\odot}$). 
There is a $\approx4\times$ decline in 
the mass normalised total SFR of clusters for galaxies with SFR
above the luminous infrared galaxies (LIRGs) SFR limit
at our redshift slice, corresponding to 2\,Gyr in look-back time. 
This is consistent with the rapid decline in
SF activity with decreasing redshift amongst cluster LIRGs
seen by previous studies using infrared-derived SFR.
\end{abstract}

\begin{keywords}
galaxies: clusters: general -- galaxies: star formation -- galaxies: evolution
\end{keywords}



\section{Introduction}
\label{sec:intro}
Probing star formation (SF) activity of galaxies in different cosmic environments paves an 
important path to 
understanding the dynamics behind galaxy evolution. Various studies have shown that SF 
decreases steadily with redshift from $z\approx2.5$ \citep[e.g.,][]{Lilly_1996,Karim2011,Sobral_2014}
and this has been shown to be the case in all environments \citep{Koyama_2013}. Results from 
\citet{Peng_2010} show that the suppression of SF activity in galaxies can be imposed by the galaxy
environment (cluster/density-dependent) and can also be a result of
internal processes (galaxy stellar mass-dependent), with both suppression effects 
acting on the galaxies independently. Some studies argue that the environment-induced
suppression is secondary to the internal/stellar mass-induced suppression
\citep[e.g.,][]{Hahn_2015}. \citet{Darvish_2016}
suggests that the environmental suppression is only relevant at
$z\leq1$ whereas the stellar mass suppression dominates at $z\geq1$. 

The internal or galaxy stellar mass regulation of SF results in a strong linear
correlation between star formation rates (SFR) and stellar mass within galaxies called 
the star formation main sequence 
\citep[MS, see][]{Brinchmann2004,Elbaz2011,Speagle2014,Schreiber2015,Katsianis2016}. 
The MS has been shown to be visible out to $z\approx6$ \citep{Speagle2014} with an
intrinsic scatter of $\approx0.2-0.3$ dex that is constant throughout cosmic time.

In the local Universe, SF activity in galaxies has been shown to be strongly dependent 
on the galaxy environment through comparative studies of galaxies in high-density (groups/clusters) environments and galaxies in the field \citep[e.g.,][]{Balogh1998,Lewis_2002, Haines_2015}.
These studies conclude that the fraction of star-forming 
galaxies in cluster environments 
is lower compared to low-density/field environments.
Galaxy clusters are populated by two broad classes of galaxies: red-and-dead elliptical 
galaxies with little to no SF activity, mostly residing in cluster cores, and blue spiral galaxies 
with ongoing SF
activity, mostly found in the cluster outskirts 
\citep[e.g.,][]{Baldry_2004, Taylor_2015, Haines_2017}. This leads to the 
relation between the fraction of star-forming galaxies and the clustercentric
radius noted by various studies 
\citep[e.g.,][]{Lewis_2002, Gomez_2003, Haines_2015}. A number of 
physical processes have been proposed and observed to occur 
within clusters leading to the suppression of SF activity among 
cluster galaxies. Some of these processes may be driven by galaxy to galaxy
interaction 
\citep[harassment or tidal interactions,][]{Farouki_1981,Moore_1996}.
In other cases, the galaxies may be sensitive to the large-scale cluster environment 
\citep[ram pressure stripping,][]{Gunn&Gott1972}
or be quenched by "burning up" as they enter 
the cluster environment \citep[][]{Larson_1980,Balogh_2000}.

Several cluster studies have observed a rapid decline in SF activity with redshift when examining the
total SFR per unit cluster mass ($\Sigma(\text{SFR})/M$). 
These studies found $\Sigma(\text{SFR})/M$ to be redshift dependent with a relation of
$\Sigma(\text{SFR})/M\propto(1+z)^{\alpha}$, with $\alpha\simeq5-7$ for clusters
out to redshifts $z\approx1.6$
\citep[see, e.g.,][]{Kodama_2004, Bai_2009, Popesso_2012, Webb_2013, Stacey_2014}.
The $\Sigma(\text{SFR})/M$ is derived from the sum of
SFR for individual galaxies identified as cluster members normalised by the cluster mass.
This quantity allows for comparison of SF activity between clusters of different masses.

SFR for normal galaxies (galaxies that do not host an active galactic nucleus) can be estimated from
various tracers at different wavelengths across the spectrum 
\citep[see, e.g., reviews by][]{Kennicutt_1998,Kennicutt&Evans2012, Madau_2014}. Rest-frame 
ultraviolet (UV) emission directly traces newly formed stars 
but suffers a great deal of attenuation from dust,
thus requiring corrections for reliable star formation estimates \citep{Bell2003}.
When the UV emission from the newly formed stars  interacts with the dust, 
it heats the dust which then re-emits at infrared wavelengths, paving a way to trace SFR via
obscured UV light \citep{Kennicutt_1998}. Emission line tracers 
(H$\alpha$, \textsc{[Oii]}, Ly$\alpha$), observed at optical/near-infrared
wavelengths are also used for SFR measurements. Just like UV emission, 
emission line tracers
face the challenge of dust attenuation and thus require
corrections to get reliable SFR estimates. Another way of tracing SFR is through
the radio continuum emission from star-forming galaxies.
Radio emission is made up of thermal free-free (bremsstrahlung) radiation and non-thermal
synchrotron radiation. The thermal free-free emission is a direct tracer of SFR
due to ionized hydrogen from massive stars in young stellar populations.
At frequencies $\leq5$\,GHz, thermal free-free radiation contributes weakly to the
radio continuum flux and non-thermal 
radiation accounts for $90\%$ of the signal \citep{Condon92}.
Due to the tight correlation between radio emission and far-infrared (FIR) emission
at 1.4\,GHz \citep[e.g.,][]{Helou_1985,Condon_1991,Yun_2001}, the radio continuum
emission has a standard model for SFR estimation at this frequency. 
Being free from dust interactions, radio emission
emerges as an advantageous tracer for estimating SFR in normal galaxies compared
to other SFR tracers (see \citet{Kennicutt&Evans2012} for a comprehensive review).

Several studies have investigated the environmental effects on the population 
and SFR of galaxies residing in high-density environments like galaxy groups and clusters.
Recently, efforts have been made toward understanding the influence of cluster 
morphology on the overall SF activity of galaxies in clusters
by comparing dynamically relaxed clusters to clusters hosting giant radio haloes and relics
(regarded as evidence of past or ongoing merger activity).
Merger/unrelaxed clusters have been observed to have
enhanced SF activity at low redshifts in individual cluster studies 
\citep[e.g.,][]{Owen_2005,Ma_2010,Ebeling_2019} as well as in statistical 
studies of multiple clusters \citep[e.g.,][]{Cohen_2014,Cohen_2015,Stroe_2017,Yoon_2020,Stroe_2021}.
Radio haloes are large-scale
extended radio emission spanning over 500\,kpc in linear size. The origin of radio haloes 
has been strongly linked with the reacceleration of relativistic particles
by cluster mergers \citep[e.g.,][]{Brunetti_2009,Cassano_2013,Lindner_2014,Kale_2015}.
Radio relics are extended synchrotron radio emission residing in the periphery
of galaxy clusters. They are elongated Mpc-scale sources linked with shock fronts
induced in the intra-cluster medium (ICM) during cluster mergers 
\citep[e.g.,][]{Ogrean_2013,vanWeeren_2016,vanWeeren_2019}.

Various processes have been suggested to occur from cluster mergers that
lead to the enhanced SF activity that is observed in merger clusters.
\citet{Bekki_2010} used numerical
simulations to show that compressed cold gas from the rising external 
pressure of ICM during mergers could explain the SF activity 
enhancement in merger clusters. \citet{Stroe_2017} suggests that the 
observed enhanced SF activity in merger clusters may be caused by shock waves in the 
ICM resulting from merging clusters or galaxy groups as well as the 
accretion of filaments. 

In this paper, we present a statistical study of
SF activity in multiple clusters using
radio observations from the MeerKAT Galaxy Cluster Legacy Survey
\citep[MGCLS DR1;][]{Knowles2021}.
MGCLS is a project of $L$-band observations (900-1670 MHz) by the MeerKAT telescope 
\citep{Jonas2016} of 115 galaxy clusters between $-80\deg$ and $15\deg$ declination.
Our MGCLS radio data is crossmatched with DECaLS photometry
to obtain a homogeneous set of radio-optical data with photo-$z$s.
With the MeerKAT field of view of $2\,$deg$^2$, we were able to probe to twice the
$R_{200}$ of our clusters in our analysis.
Extended radio emission is detected in 15 out of 20 clusters in our sample,
5 of which are new detections, and marked
as candidates for radio haloes, relics and a mini-halo depending on
their morphology \citep[see][]{Knowles2021}. Fourteen clusters in our sample host
extended emission in the form of either a radio-halo, a relic or a phoenix
\citep[a subclass of a relic,][]{vanWeeren_2019}.
With this article, we aim to investigate SF activity in cluster environments
and its relationship with the dynamical state of the cluster.
We compare our results to previous studies of multiple clusters
carried out using FIR as a tracer for SFR.

The structure of this paper is as follows.
In Section \ref{sec:data} we describe the radio data, photometric
data, cluster masses as well as the AGN identification and membership selection process.
In Section
\ref{sec:sfr} we discuss our SFR estimates, fraction of star-forming galaxies, 
SFR and cluster mass relation as well as the 
mass normalised total SFR relation with redshift. 
We discuss our results in Section \ref{sec:discussion}. We end the paper by
summarising our conclusions and laying out some future work in Section \ref{sec:conclusion}.

We assume a cosmology with $H_{0}$ = 70\,km\,s$^{-1}$\,Mpc$^{-1}$, $\Omega_{\rm m,0} = 0.3$ and
$\Omega_{\rm \Lambda,0} = 0.7$. $M_{200}$ denotes the mass enclosed within a sphere of radius $R_{200}$
in which the mean enclosed overdensity is equal to 200 times the critical density of the Universe 
at the cluster redshift. The \citet{Chabrier_2003} initial mass function (IMF) is used for
SFR calculations.

\section{Data}
\label{sec:data}

\subsection{Cluster sample}
\label{sec:sample}
The selection of the 20 clusters from the parent MGCLS sample was guided by the availability of 
cluster photometric data in the Dark Energy Camera Legacy Survey \citep[DECaLS DR8;][]{Dey_2019}
as well as the availability of cluster masses\,($M_{200}$) from the Atacama 
Cosmology Telescope \citep[ACT DR5;][]{Hilton_2021}. The ACT DR5 catalogue
has a homogeneous selection of clusters with consistently derived mass estimates 
which consists of 4195 optically confirmed Sunyaev-Zel'dovich 
(SZ) galaxy clusters with redshift measurements. The ACT DR5 
catalogue provides a set of mass estimates that have been re-scaled according to a 
richness-based weak-lensing mass calibration through a technique described
in \citet{Hilton_2018}. The DECaLS photometric data is complete for WISE (3.4\,\micron{}) W1 band apparent magnitude (mag) $m_{3.4}\leq19.7$. This corresponds to an absolute mag limit $M_{3.4}=-20.8$, K-corrected to our sample median redshift ($z=0.25$). We are sensitive down to a limit of $\approx M\substack{*\\3.4}$+1.5, using $m^{*}$ (18.2) of the IR-selected cluster sample luminosity function best-fit from the Spitzer Infrared Array Camera (IRAC) $3.6\,\micron{}$-band \citep{Mancone_2010}. $M^{*}$ is scaled down by 0.08 mag to correct for the colour difference between W1 (3.4\,\micron{}) and Spitzer IRAC channel one (3.6\,\micron{}). The IR luminosity function is a good tracer of the stellar mass function since the 3.6\,\micron{} photometry probes the peak of the stellar light.

The MGCLS catalogue is made up of a heterogeneous sample of 115 clusters with no 
general selection criteria guided by cluster mass or redshift. MGCLS comprises
two subsamples grouped as ``radio-selected'' and ``X-ray-selected'' with each
subsample contributing 41 and 74 clusters respectively. 
The MGCLS radio-selected subsample was picked to be biased towards 
high-mass clusters with extended radio emission. The X-ray-selected subsample
was guided by the Meta-Catalogue of X-ray-detected Clusters \citep[MCXC;][]{MCXC_2011},
which is a heterogeneous compilation of X-ray-selected clusters
with no direct prior biases towards or against clusters with extended radio emission.

Of the total 115 MGCLS clusters, 66 had optical counterparts with complete coverage in DECaLS 
and 38 of the 66 clusters have cluster masses from the ACT catalogue. 
The MGCLS field data is classified by image data quality ranging from 0 to 3. The data quality from 
0 to 1 is classified as good to moderate dynamic range whereas 2 to 3 is classified as poor dynamic range
with ripples and source artefacts on the data. For this study, we only work with fields that are
flagged 0 and 1 to ensure that our data has the least amount of contamination. 

\begin{figure}
\includegraphics[width=\columnwidth]{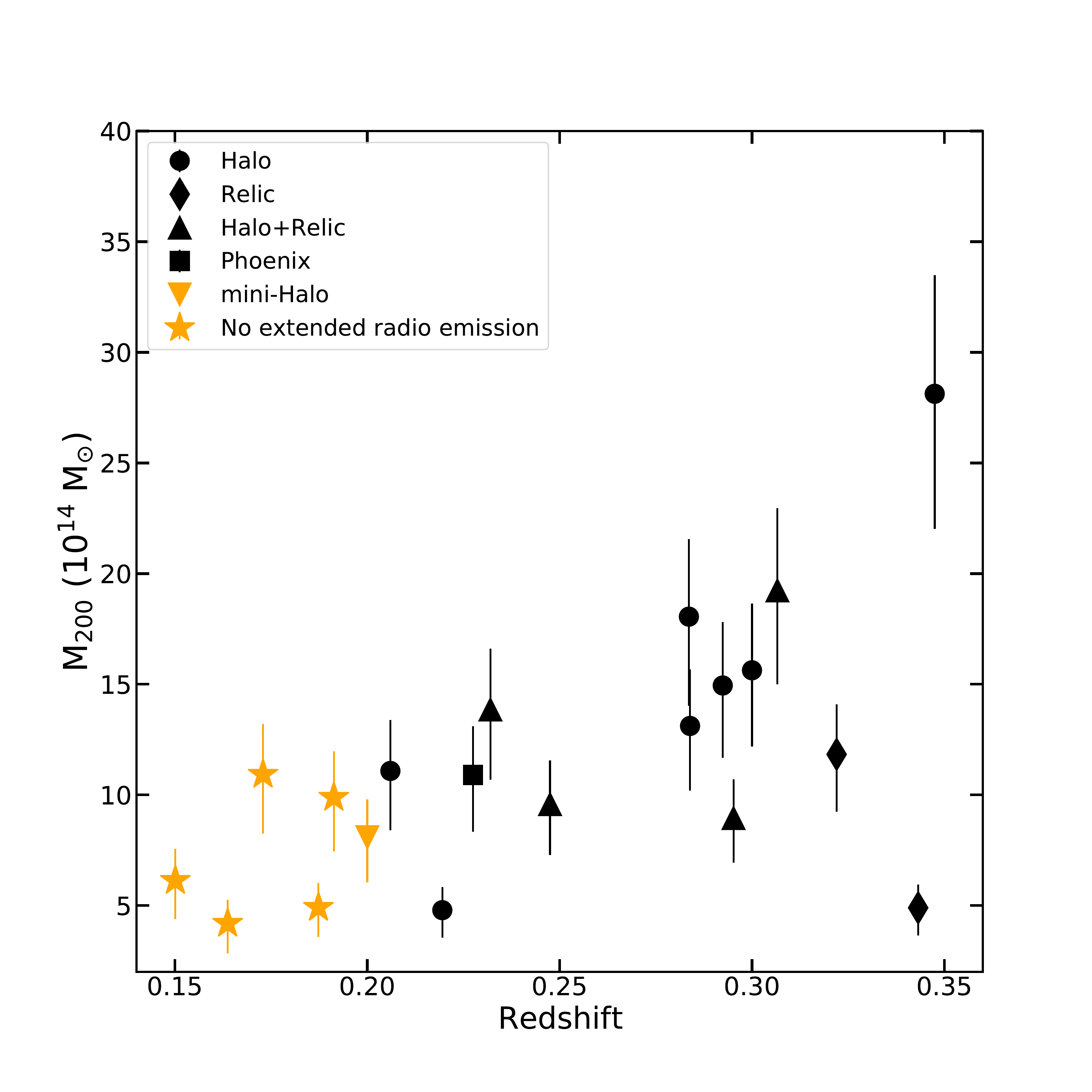}
\caption{The mass ($M_{200}$) and redshift distribution of the
cluster sample used in this work. We use a subset of the SZ-selected 
ACT DR5 sample \citep{Hilton_2021} to ensure a homogeneous
set of cluster mass estimates. The black markers show clusters with merger-linked extended radio emission and the orange markers show relaxed clusters with no extended emission and non-merger-linked extended emission. Each symbol represents the type of extended radio emission a cluster hosts.}
\label{fig:z_bins}
\end{figure}

Figure \ref{fig:z_bins} shows the mass distribution of our final cluster
sample with redshifts for 20 clusters which have reliable 
photo-$z$s estimates and ACT SZ-derived cluster masses, all falling in 
the redshift range $0.15<z<0.35$. The cluster sample selection was limited to
clusters with $z>0.15$ due to the large scatter observed in
photo-$z$s estimates at $z<0.15$, making them unreliable
(see Figure \ref{fig:photozComparison}). The cluster sample properties are listed
in Table \ref{tab:clusters_}. Our final sample has 10 clusters from the
radio-selected subsample, 9 of which are hosts of extended emission in the form of a 
radio halo and/or relic and 1 cluster with no extended emission. 
The other 10 clusters in our final sample are from the X-ray-selected
subsample with 5 hosting extended emission in the form of a radio halo and/or relic,
1 hosting extended emission in the form of a mini-halo and 4 with no extended emission.

Fourteen clusters in our sample host large-scale
extended emission (haloes and relics) linked to on-going merger activity. The presence
of large-scale extended emission has been linked
to the dynamical state of mergers or unrelaxed clusters by 
\citep[e.g.,][from their study of haloes using radio data and X-ray observations]{Cassano_2010}.
The 6 clusters
with no large-scale extended emission are assumed to be relatively dynamically relaxed or 
undergoing minor mergers and likely to be less disturbed 
in contrast to the 14 clusters hosting haloes and relics. 
We conducted a visual inspection on the dynamical state of 4 of the 6 non-halo/relic-hosting
clusters in our sample using archival X-ray imaging from \textit{Chandra} and \textit{XMM Newton} surveys.
The X-ray images indicate that the clusters are likely
to be dynamically relaxed. \citet{Lovisari_2017} classified the dynamical state of some of
the clusters that overlap with our sample as either `disturbed', `relaxed', or `mixed', 
using criteria based on seven parameters. Clusters J0449.9-4440 and
J0525.8-4715 are classified as relaxed systems while J0510.2-4519 (hosting a mini-halo) is noted as
a mixed system by the \citet{Lovisari_2017} classification criteria. They 
point out that the mixed systems classification may be subjective and that
a cluster falling under this category depends on the parameters chosen to optimise
completeness and purity.
Given the high masses of clusters in our sample, the clusters without large-scale extended 
emission are likely relaxed systems or have only minor mergers and can be considered to be
less disturbed environments for star-forming galaxies.

\begin{table*}
	\centering\makegapedcells
	\begin{tabular}{lccccccccc}
\hline
\hline
\hspace{-1.5cm}(1)&(2)&(3)&(4)&(5)&(6)&(7)&(8)&(9)&(10)\\
\textbf{Cluster} &         \textbf{RA} &        \textbf{DEC} &      \textbf{z} &  \textbf{RMS} & \textbf{$M_{200}$ } &  \textbf{$R_{200}$} & \textbf{$f_{\textbf{SF}}$} & \textbf{$\Sigma{\textbf{SFR}}$} & \textbf{State} \\
\hspace{-1.5cm}
\textbf{} &         \textbf{} &        \textbf{} &      \textbf{} &  \textbf{($\mu$J b$^{-1}$)} &  \textbf{(10$^{14}$ M$_{\odot}$)} &  \textbf{(Mpc)} & \textbf{} & \textbf{M$_{\odot}$yr$^{-1}$}  & \textbf{}\\
\hline
        Abell 209 &  22.970833 & -13.609444 & 0.206 &  3.6   &  $11.1 \substack{+2.7 \\ -2.3}$   & 2.0    & $0.242$ &  $380.54\pm 42.87  $   &    halo\\
       Abell 2744 &   3.578333 & -30.383333 & 0.307 &  2.9   &  $19.2 \substack{+4.2 \\ -3.7}$   & 2.3    & $0.113$ &  $366.31\pm 68.29   $   &    halo, relic\\
       Abell 2813 &  10.851667 & -20.621389 & 0.292 &  3.4   &  $15.0 \substack{+3.3 \\ -2.9}$   & 2.1    & $0.084$ &  $272.05\pm 50.64  $   &    halo$^{\ddag}$\\
       Abell 2895 &  19.546250 & -26.973056 & 0.228 &  3.0   &  $11.0 \substack{+2.6 \\ -2.2}$   & 2.0    & $0.241$ &  $302.31\pm 32.77   $   &    phoenix$^{\ddag}$\\
        Abell 521 &  73.537917 & -10.238611 & 0.248 &  3.4   &  $9.6  \substack{+2.3 \\ -2.0}$  & 1.9     & $0.132$ &  $309.21\pm 45.54   $   &    halo, relic\\
       Abell S1063& 342.181250 & -44.528889 & 0.348 &  2.6   &  $28.1 \substack{+6.1 \\ -5.4}$  & 2.6     & $0.194$ &  $682.19\pm 109.97  $   &    halo\\
       Abell S295 &  41.399167 & -53.038056 & 0.300 &  2.3   &  $15.6 \substack{+3.4 \\ -3.0}$  & 2.2     & $0.149$ &  $458.36\pm 65.78  $   &    halo\\
     J0051.1-4833 &  12.796667 & -48.559722 & 0.187 &  2.6   &  $4.9  \substack{+1.3 \\ -1.1}$ & 1.5      & $0.133$ &  $106.24\pm  6.36   $   &   relaxed\\
     J0217.2-5244 &  34.302500 & -52.746944 & 0.343 &  2.8   &  $4.9  \substack{+1.2 \\ -1.1}$ & 1.4      & $0.146$ &  $245.21 \pm  47.80 $   &   relic$^{\ddag}$\\
     J0225.9-4154 &  36.477500 & -41.909722 & 0.220 &  2.7   &  $4.8  \substack{+1.2 \\ -1.0}$ & 1.5      & $0.139$ &  $103.29\pm  15.59 $   &   halo\\
     J0232.2-4420 &  38.070000 & -44.347500 & 0.284 &  2.6   &  $18.1 \substack{+4.0 \\ -3.5}$  & 2.3     & $0.229$ &  $452.7 \pm   57.25 $   &    halo\\
     J0336.3-4037 &  54.077917 & -40.622222 & 0.172 &  3.5   &  $10.9 \substack{+2.7 \\ -2.3}$  & 2.0     & $0.12$ &   $161.97 \pm   12.07 $   &    relaxed\\
     J0449.9-4440 &  72.480000 & -44.678056 & 0.150 &  2.6   &  $6.1  \substack{+1.7 \\ -1.4}$ & 1.7      & $0.074$ &  $89.59 \pm   3.58 $   &   relaxed\\
     J0510.2-4519 &  77.557500 & -45.321111 & 0.200 &  3.0   &  $8.1  \substack{+2.0 \\ -1.7}$ & 1.8      & $0.123$ &  $172.26\pm 10.28  $   &   relaxed, mini-halo$^{\ddag}$\\
     J0516.6-5430 &  79.158333 & -54.514167 & 0.295 &  3.1   &  $9.0  \substack{+2.0 \\ -1.8}$ & 1.8      & $0.108$ &  $241.93\pm 36.16 $   &   halo,relic\\
     J0525.8-4715 &  81.465000 & -47.250556 & 0.191 &  3.0   &  $9.9  \substack{+2.4 \\ -2.1}$ & 1.9      & $0.173$ &  $242.16\pm 30.04  $   &   relaxed\\
     J2023.4-5535 & 305.850000 & -55.591700 & 0.232 &  2.7   &  $13.9 \substack{+3.2 \\ -2.8}$  & 2.1     & $0.201$ &  $360.01\pm 42.34  $   &    halo, relic\\
MACS J0257.6-2209 &  44.422083 & -22.153889 & 0.322 &  3.2   &  $11.8 \substack{+2.6 \\ -2.3}$  & 2.0     & $0.114$ &  $310.40\pm 49.84 $   &    relic$^{\ddag}$\\
RXC J0528.9-3927  &  82.234583 & -39.462778 & 0.284 &  2.6   &  $13.1 \substack{+2.9 \\ -2.6}$  & 2.1     & $0.193$ &  $311.16\pm 46.45  $   &    halo\\
RXC J0543.4-4430  &  85.851667 & -44.505278 & 0.164 &  3.6   &  $4.2  \substack{+1.3 \\ -1.1}$  & 1.5     & $0.151$ &  $135.02\pm 12.59 $   &    relaxed\\
\hline
\end{tabular}
\caption{The cluster sample. Columns: (1) Name of the cluster; (2,3) MeerKAT pointing coordinates: 
J2000 Right Ascension and Declination in degrees; (4) Cluster redshift; (5) MeerKAT image sigma-clipped 
standard deviation in micro-Jy per beam; 
(6) Cluster mass in $10^{14}$M$_{\odot}$; (7) virial radius $R_{200}$ in Mpc; 
(8) The fraction of all star-forming galaxies 
within $R_{200}$ (see Section \ref{sec:fsf_r200}); (9) Total SFR for galaxies within $R_{200}$
(see Section \ref{sec:totSFR_M}); (10) The dynamical state of a cluster.
The clusters with newly detected
extended radio emission in the MGCLS are marked as candidates based on their morphology
are indicated by $^{\ddag}$.}
\label{tab:clusters_}
\end{table*}

\renewcommand*{\thefootnote}{\arabic{footnote}}
\subsection{MeerKAT data}
\label{sec:mkat}
The MGCLS DR1 \citep{Knowles2021}
is a publicly released catalogue of 115 galaxy clusters observed by the 
MeerKAT telescope in the $L$-band throughout 900 – 1670 MHz. Each cluster is observed at 
full polarization for $\approx$ 6 – 10 hours. The MGCLS DR1 offers raw MeerKAT 
continuum visibilities, basic image cubes at $\approx 8''$ resolution as well as
enhanced spectral and polarization image cubes at $\approx$ $8''$ and $15''$ 
resolutions. MGCLS image products are sensitive within a range of
$\approx$ 3\,$\mu$Jy beam$^{-1}$ for an $8''$ beam to 
$\approx$ 10\,$\mu$Jy beam$^{-1}$ for a $15''$ beam.
The basic cubes span a $2\,$deg$^2$ field of view (FoV).
The enhanced data products are primary beam corrected within a $1.2\,$deg$^2$ FoV.
The survey gives good detections up to $\approx$ $10'$ and has
a wide bandwidth that enables spectral and Faraday rotation mapping.

The Python Blob Detection and Source Finder \citep[\texttt{PyBDSF};][]{pybdsf2015} software was
used to create catalogues for all 115 clusters from the 1.28 GHz images
at a threshold of $5\sigma$ source detection 
\citep[see][for more details on source detection]{Knowles2021}.
The complete MGCLS DR1 catalogue has a combined total of $\approx 626000$ compact sources 
from all 115 cluster data sets.

\subsection{DECaLS data}
\label{sec:decals}
The DECaLS data comes from images captured by the Dark Energy Camera
\citep[DECam $\approx3.2\,$deg$^2$ FoV;][]{Flaugher_2015} located at the 4-m Blanco 
telescope at the Cerro Tololo Inter-American Observatory. DECam is a highly
sensitive camera optimised for a wide-field survey across a broad wavelength
range ($\approx400-1000$\,nm). DECaLS is part of the three surveys that integrate 
to make the DESI Legacy Survey,
alongside the Beijing-Arizona Sky Survey \citep[BASS;][]{Zou_2017} and the 
Mayall z-band Legacy Survey \citep[MzLS;][]{MzLS_2016}.
DECaLS provides optical $grz$ photometry combined with 3.4, 
4.6\,\micron{} photometry from the Wide-field Infrared Survey 
Explorer mission \citep[WISE;][]{Wright_2010}. The DECaLS $grz$ photometry
data has 5$\sigma$ depth levels of $g=24.0$, $r=23.4$ and $z=22.5$ in AB magnitudes.

\subsection{Photometric redshifts}
\label{sec:zphots}

\begin{figure}
\includegraphics[width=\columnwidth]{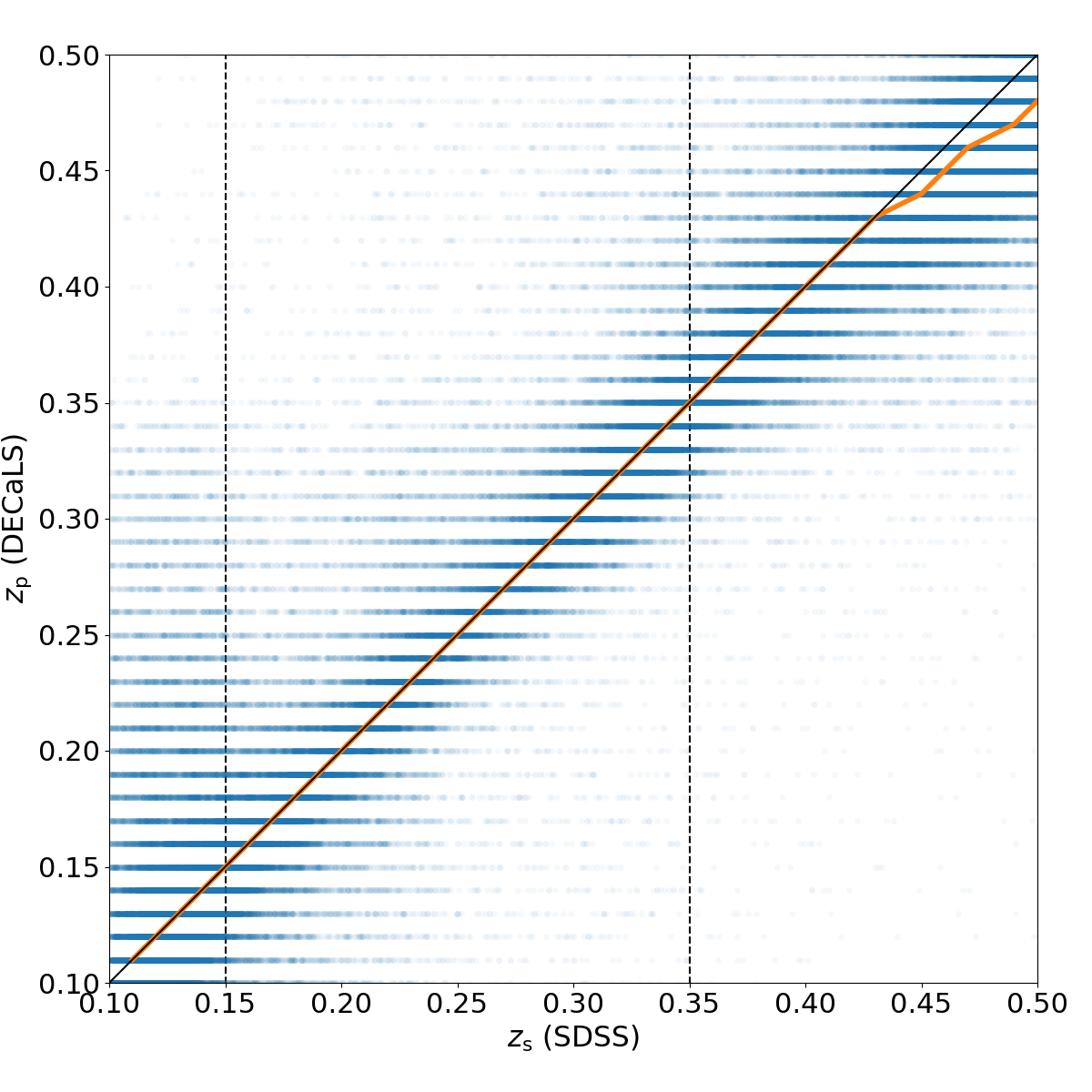}
\caption{Comparison of photo-$z$s 
estimated using \texttt{zCluster} on DECaLS photometry with spec-$z$s 
from SDSS (219,380 galaxies, $0.1<z<1$). The photo-$z$s are evaluated on a grid 
with step size $\Delta z = 0.01$, 
hence the linear order of data points along the vertical axis. The orange line 
shows the median redshift residual in bins of width 
$\Delta z_{\rm s} = 0.2$. It becomes slightly
biased at higher redshift, but this is well above 
our cluster sample redshift range as indicated by the vertical dashed lines. The photo-$z$s
do not perform well at $z_{\rm s} < 0.15$ due to the lack of $u$-band
mags in DECaLS.}
\label{fig:photozComparison}
\end{figure}

We use the \texttt{zCluster}\footnote{\url{https://github.com/ACTCollaboration/zCluster}} package 
\citep[see,][]{Hilton_2021, Pillay_2021} to 
estimate the maximum likelihood 
galaxy photo-$z$s and their 
probability distribution, using a 
template-fitting method. We use a combination of the 
\citet{ColemanWuWeedman_1980} templates and a 
subset of the spectral templates used in the COSMOS
survey \citep{Ilbert_2009, Salvato_2011}, 
representing a range of normal galaxies and AGNs. 
We test the accuracy of the photo-$z$s
using 219,380 galaxies within the DECaLS footprint 
that have spec-$z$s in SDSS DR16 
\citep{2020ApJS..249....3A}. We also use this sample to
calibrate zero-point offsets to the DECaLS photometry
to minimise the bias of the photometric redshift (photo-$z$)
residuals ($\Delta z = z_{\rm s} - z_{\rm p}$, 
where $z_{\rm s}$ is the spectroscopic redshift (spec-$z$), and
$z_{\rm p}$ is the photo-$z$). After 
this procedure, we find mean 
$\Delta z/(1+z_{\rm s}) = 0.008$, and scatter 
$\sigma_{\rm bw} = 0.03$, as estimated using
the biweight scale \citep[e.g.,][]{Beers_1990}. 
We find that 93\% of the galaxies have 
$\Delta z/(1+z_{\rm s}) < 0.09$, i.e., 7\% of the 
photo-$z$s are expected to be 
``catastrophic outliers''. Figure~\ref{fig:photozComparison}
shows a comparison of the photometric and 
spec-$z$s for this test sample.

\subsection{Cluster membership selection}
\label{sec:members}

The most unequivocal way to select cluster members
is by use of reliable spec-$z$s. However,
extensive spec-$z$ catalogues covering 
our cluster sample are not available.  
For this study, we rely solely on photo-$z$s 
to achieve a homogeneous dataset. In Section \ref{sec:zphots}, 
we discuss the process for deducing the photo-$z$ using
\texttt{zCluster} and their accuracy.

Instead of relying only on a single best-fitting photometric redshift estimate for each galaxy without considering its uncertainty, our membership process uses the full redshift probability distribution $P(z)$ provided by \texttt{zCluster}.
We follow a method developed by \citet{Pello_2009} to identify cluster members. 
This method builds on a technique by \citet{Brunner_2000}, 
which calculates the probability ($P_{\rm member}$) of being a member galaxy at a redshift range centred 
at the cluster redshift ($z_{\text{cl}}$), with a width ($\delta z$) that 
depends on the accuracy of the photo-$z$s (see Section \ref{sec:zphots}),

\begin{equation}
    P_{\rm member} = \int_{z_{\text{cl}}-\delta z}^{{z_{\text{cl}}+\delta z}} P(z) \,dz \,.
\label{eq:1}
\end{equation}

For our case, $\delta z = n\sigma_{\rm bw}(1+z_{\rm cl})$. 
This approach works well for this study as we mainly focus on the integrated 
properties of the cluster galaxy population (i.e., total cluster SFR or 
fraction of star-forming galaxies), rather than the properties of individual galaxies.
We use the three clusters in our sample with available extensive spec-$z$s (A209, A2744, AS1063) in the literature to calibrate cluster membership selection for the rest of the sample. We classify galaxies with spec-$z$s as members that satisfy the condition $|z_{\rm cl}-z_{\rm spec}|<3\sigma_{\rm cl}(1+z_{\rm cl})$, where $\sigma_{\rm cl}$ is the cluster velocity dispersion. The Abell 209 spec-$z$s were obtained from 
the Cluster Lensing And Supernova survey with Hubble \citep[CLASH-VLT;][]{Annunziatella_2016} and the Arizona Cluster Redshift Survey \citep[ACReS; described in][]{Haines_2015}. The spec-$z$s for Abell 2744 are from the Multi Unit Spectroscopic Explorer \citep[MUSE;][]{MUSE_2018} and Anglo-Australian Telescope \citep[AAT;][]{AAOmega_2011} catalogues. The Abell S1063 spec-$z$s were obtained from the CLASH-VLT catalogue \citep{Mercurio_2021}.

From the stacked catalogue of all the three clusters with spec-$z$s, we get $\sigma_{\rm bw}=0.08$ from the redshift residuals, $\Delta z/(1+z_{\rm s})$. The spectroscopic completeness level $C=82\%$ and the field contamination $F=17\%$ using $n=2$ and $P_{\rm member}\geq0.5$. The $n$ and $P_{\rm member}$ values were chosen to maximise $C$ and minimise $F$ while selecting galaxies that are likely to be members using redshift probability distribution functions. The spectroscopic completeness level describes the number of confirmed spec-$z$ members correctly identified as members by the photo-$z$s selection criteria. The field contamination shows the percentage level of spec-$z$s that are non-members/field galaxies but selected by the photo-$z$s membership criteria.

The $\sigma_{\rm bw}$ from the stacked catalogue is higher than the one estimated with the SDSS spec-$z$s mentioned in section \ref{sec:zphots}. We adopt the larger value as a more conservative estimate of the cluster member photometric redshift accuracy. Table \ref{tab:zSpec_clusters} shows the $C$ and $F$ percentage levels reached for each cluster. We use the same selection parameters to identify members using only photo-$z$s for all our 20 clusters, assuming a $C>80\%$ and $F<20\%$ for our full sample. We further attempt to mitigate the field contamination in two ways: (1) we check the effect of using a higher $P_{\rm member}$ on our results, and (2) we use $P_{\rm member}$ as a weight for each galaxy when investigating integrated cluster properties. We find that increasing the $P_{\rm member}$ threshold has minimal effect on our overall conclusions while reducing the completeness level in the spectroscopic sample and with little change in the field contamination level.

We obtain a total of 14,419 galaxies out to 2$R_{200}$ for the W1 complete DECaLS membership sample for the 20 clusters. This gave us a sample of 3054 cluster member galaxies observed by MeerKAT. The MeerKAT-detected member galaxies sample is complete above a $5\sigma$ detection limit of SFR = 1.9\,M$_{\odot}\,$yr$^{-1}$ in our most distant cluster ($z=0.35$). 


\begin{figure*}
\centering
\vspace{-0.5cm}
\includegraphics[trim=11.25cm 3.5cm 2.5cm 0.cm,scale=.8]{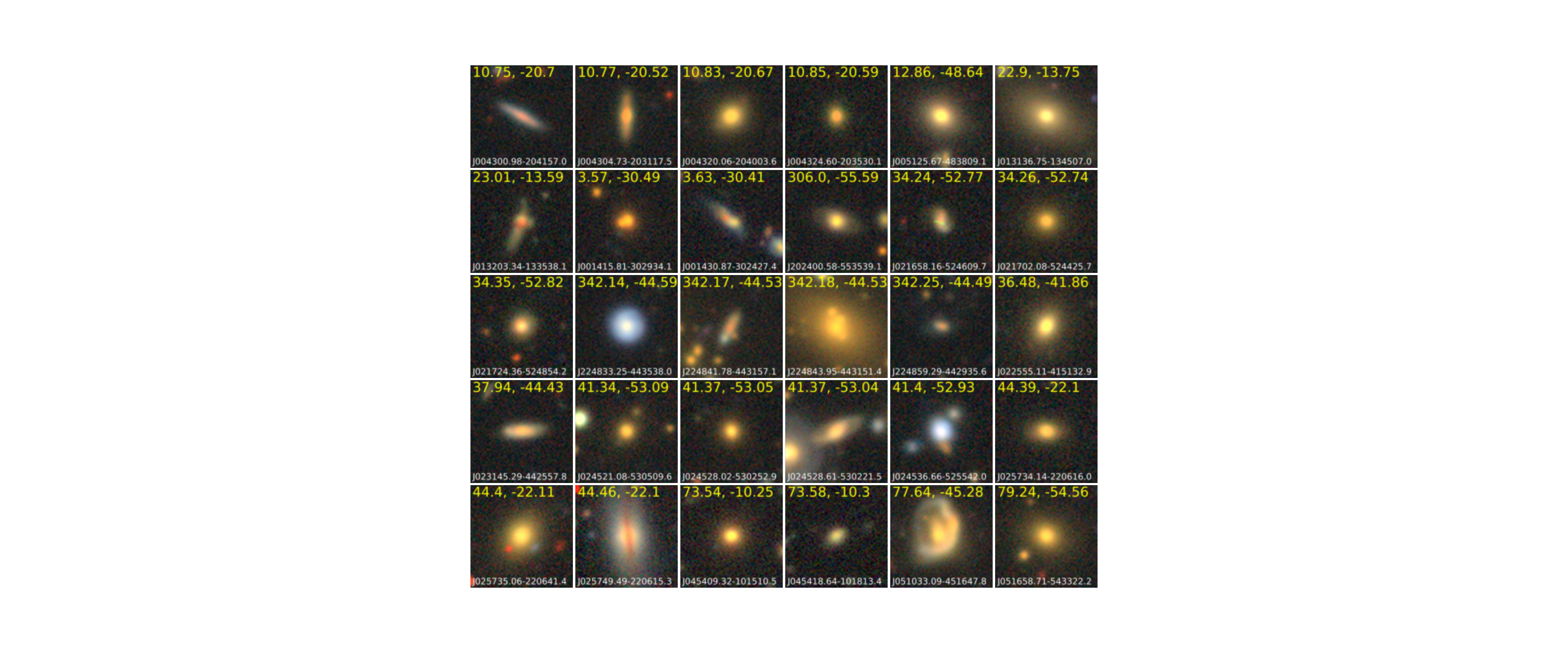}
\vspace{1.5cm}
\caption{Thumbnail images of galaxies within $R_{\rm 200}$ classified as radio-loud AGN.
Each thumbnail has the galaxy identified as radio-loud centred on the image, the
coordinates of the galaxy at the top of the image and the MeerKAT source
ID at the bottom.}
\label{fig:radio_loud}
\end{figure*}

\begin{table}
	\centering\makegapedcells
	\begin{tabular}{lccccc}
\hline
\hline
(1)&(2)&(3)&(4)&(5)&(6)\\
\textbf{Cluster} & \textbf{$z_{\rm cl}$} & \textbf{No. of spec-$z$s} & \textbf{$\sigma_{\rm bw}$} & \textbf{$C$} (\%) & \textbf{$F$} (\%)\\
\hline
Abell 209   & 0.206 & 2377 (753) & 0.10 & 81 & 15\\
Abell 2744  & 0.307 & 971 (463) & 0.06 & 81 & 13\\
Abell S1063 & 0.348 & 1626 (751) & 0.08 & 84 & 25\\
\hline
\end{tabular}
\caption{Quantities derived from the three clusters in our sample with spec-$z$s: (1) Cluster name; (2) Cluster redshift; (3) the total number of matched spectroscopic galaxies within $R_{200}$. The number of spec-$z$ galaxies within $|z_{\rm cl}-z_{\rm spec}|<3\sigma_{\rm cl}(1+z_{\rm cl})$ is shown in brackets; (4) the scatter in the redshift residual derived from the biweight estimator; (5) spec-$z$s completeness level; (6) fraction of spec-$z$ field galaxies.}
\label{tab:zSpec_clusters}
\end{table}


\subsection{AGN identification}
\label{sec:agn}

Galaxies hosting active galactic nuclei (AGN)
pose a significant challenge when using radio observations to 
estimate SFR in galaxies. 
Previous studies \citep[e.g.,][]{Sadler2002,Condon2002,Mauch_2007}
have shown that radio-AGN
may contribute up to $50\%$ of the radio
luminosities in the local Universe
at luminosities just below $L_{\rm 1.4GHz}
\approx 10^{23}$ W\,Hz$^{-1}$, and
that at luminosities over $10^{23}$ W\,Hz$^{-1}$ AGN
dominate the population over star-forming 
galaxies. It is therefore essential for us
to remove AGN-hosting galaxies from our 
dataset to achieve an unbiased star-forming sample. To identify X-ray AGN, we cross-matched our
member galaxies sample with the \textit{Chandra} serendipitous source catalogue
\citep[CSC 2.0;][]{Evans_2019} and the fourth \textit{XMM-Newton} serendipitous 
source catalogue \citep[4XMM-DR11;][]{Webb_2020}. 
X-ray sources above the rest-frame (2-10 keV) X-ray 
luminosity, $L_{\text{X}} > 10^{42}$\,erg\,s$^{-1}$ are expected 
to be AGN, whereas those below this limit are expected to be powered by star formation. We used
the cross-matching radius of $4''$ for both X-ray catalogues, yielding
65 and 54 sources respectively. There are four clusters in our sample with
no coverage or matches within the X-ray catalogues.
Of the 119 galaxies with X-ray 
cross-matches, we remove 94 galaxies that are above the X-ray 
AGN luminosity cut. 

To determine additional AGN, we adopt the `R90' WISE IR-selection criteria by \citet{Assef_2018}
\citep[see also][]{Stern_2012,Assef_2013} for
AGN classification. This uses only the WISE W1 and W2 bands to 
identify AGN based on their W1-W2 colour, 
compared to a threshold that depends 
on the W2 band mag 
\citep[see equation 4 of][which identifies AGNs with 90\% reliability]{Assef_2018}.
With this AGN separation method, we removed 
28 sources classified as AGN-hosting galaxies from our sample.

\begin{figure*}
    \includegraphics[width=0.331\textwidth]{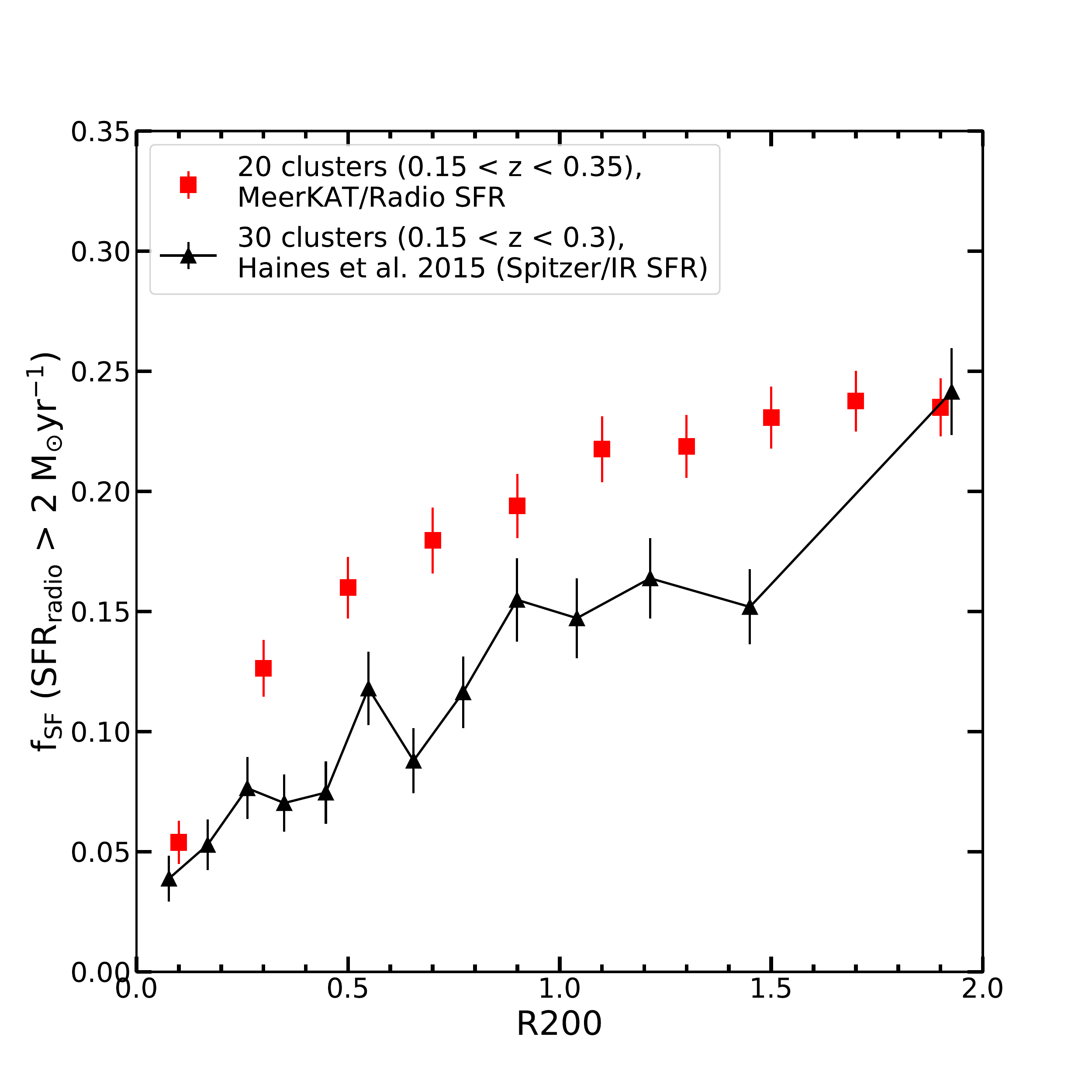} 
    \includegraphics[width=0.331\textwidth]{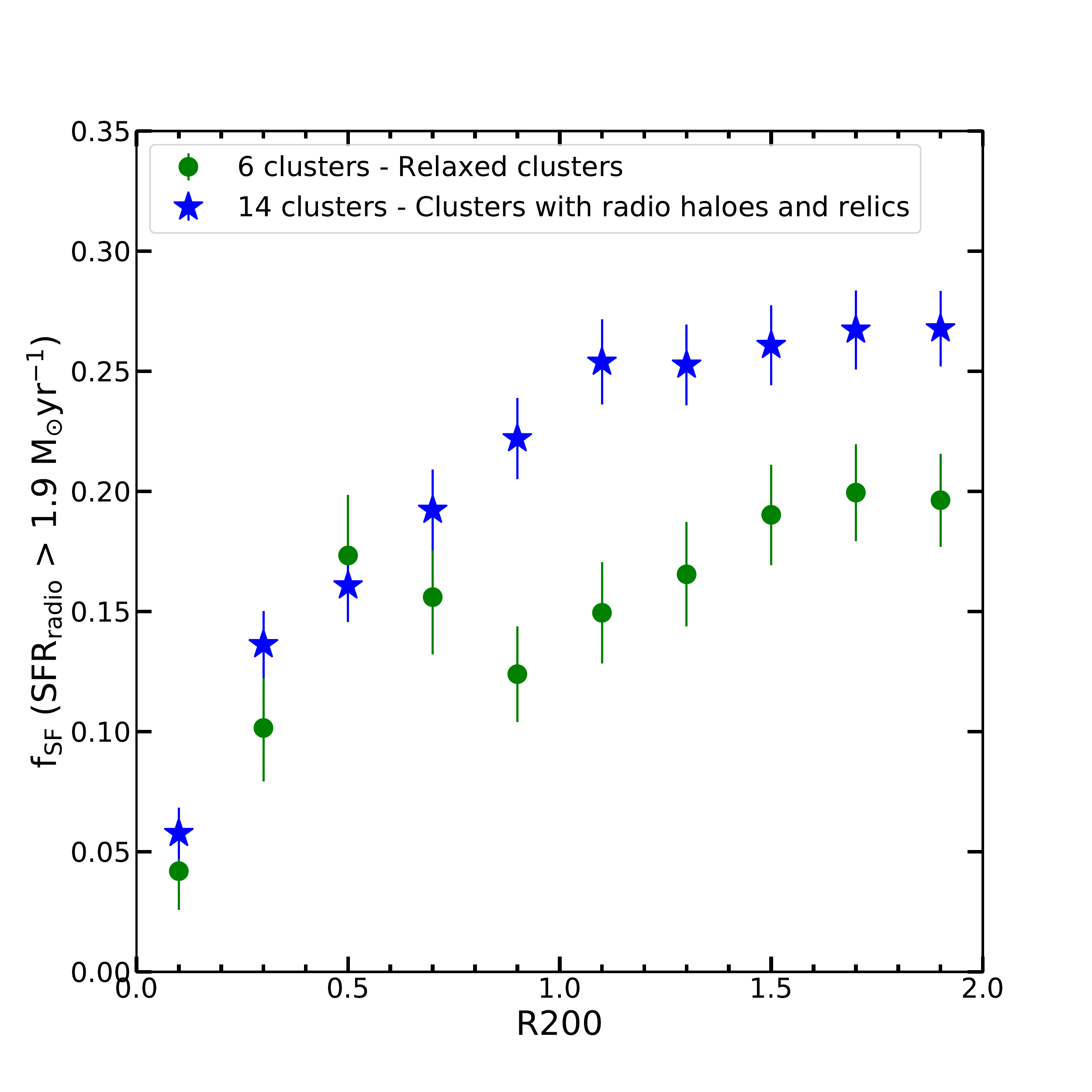} 
    \includegraphics[width=0.331\textwidth]{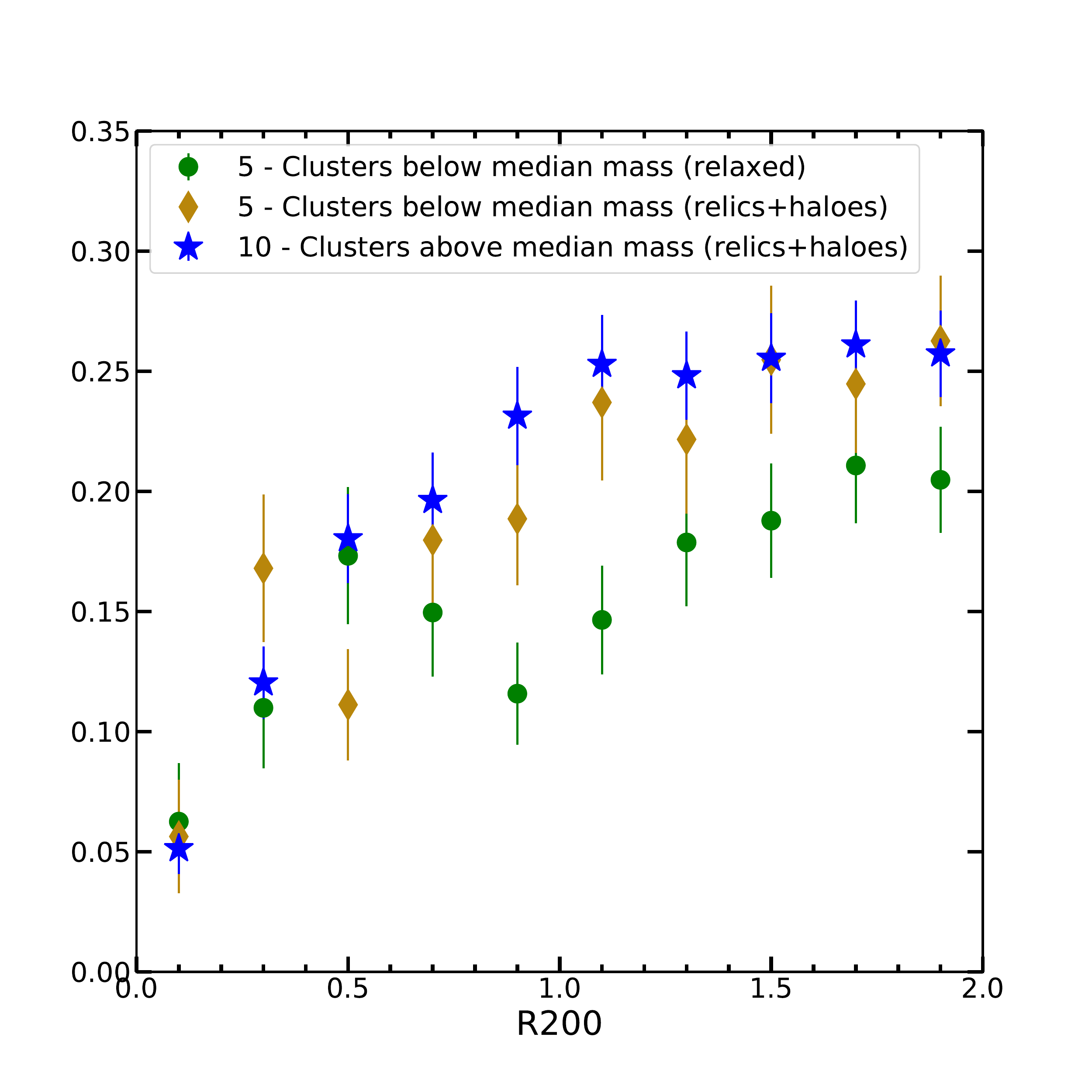}
        
\caption{\textit{Left panel}: The fraction of star-forming galaxies, $f_{\rm SF}$ as a
function of projected radial distance, in units of $R_{200}$ from the cluster
centre. We only plot MGCLS cluster members with SFR greater 
than the $5\sigma$ detection limit of 
SFR = 2 M$_{\odot}\text{yr}^{-1}$ for comparison with IR-derived results from
\citet{Haines_2015} using \textit{Spitzer}/MIPS 24\,$\mu$m fluxes down to the same SFR limit.
The red squares indicate $f_{\rm SF}$ values for MeerKAT-detected cluster members in 20 clusters
at $0.15<z<0.35$ and their 1$\sigma$ uncertainties.
The black triangles show the $f_{\rm SF}$ values by \citet{Haines_2015} in 30
clusters at $0.15<z<0.30$ with binomial statistics derived uncertainties.
\textit{Middle panel}: The comparison of $f_{\rm SF}$ trends between clusters 
hosting radio haloes/relics (blue star markers) and relaxed clusters (green circles). 
\textit{Right panel}: The $f_{\rm SF}$ trends of clusters split
by median cluster mass ($1.09\times10^{15}\,M_{\odot}$) and by cluster dynamical state.
}
\label{fig:frac_haines}
\end{figure*}

Finally, to reduce contamination from radio-loud AGN, we apply a radio luminosity cut on
our sample for galaxies  with $L_{\rm 1.4GHz} > 10^{23}$ W\,Hz$^{-1}$ \citep{Condon2002}. 
This cut removes 80 galaxies, some of which
may be  star-forming, but a relatively
small number (<3$\%$) of the total sample. 
Only 30 galaxies with the luminosity above
the radio-loud AGN cut are within our total sample $R_{200}$.
Figure \ref{fig:radio_loud} shows Legacy Surveys DR9 images of the 30
radio-loud galaxies within $R_{200}$ 
of our cluster sample to visually confirm AGN status.
Most of the galaxies show evidence of early-type morphology and are
likely to be AGN hosts and not star-forming. A handful of galaxies
show disky morphology with observed dust lines visible across their images.
We observe three galaxies within the radio-loud sample with ambiguous morphology.
These are; MKTCS J024536.66-525542.0 and MKTCS J224833.25-443538.0, which
shows the structure that is consistent with
quasar morphology and the MKTCS J051033.09-451647.8, which 
appears to be two galaxies colliding with two noticeable bulges
and disturbed spiral arms.
It is also worth noting that low 
luminosity radio-loud AGN may be left out by the radio luminosity cut and
we assume that their number does not significantly affect our results.

After removing 203 galaxies identified as
AGN hosts, we obtain a total sample of 2851 MeerKAT-detected galaxies
out to 2$R_{200}$, which we assume to be powered by star 
formation in the sections that follow. After removing 436 X-ray and WISE AGN from the photometric DECaLS selected members sample, we obtain 13,983 galaxies.

\section{Star formation rates}
\label{sec:sfr}

We used the \citet[][]{Bell2003} calibration of the radio--FIR
correlation \citep[see also][]{Karim2011} to convert radio luminosities to SFR
scaled down by 1.74 from the Salpeter IMF to the Chabrier IMF:

\begin{equation}
    {\rm SFR}~(\rm M_{\odot}\,{\rm y}r^{-1}) = 
        \begin{cases}
            3.18\times10^{-22}L, & L>L_{\rm c}\\
            \dfrac{3.18\times10^{-22}L}{0.1+0.9(L/L_{\rm c})^{0.3}}, & L\leq L_{\rm c}
        
        \end{cases}
\end{equation}
where $L = L_{\rm 1.4GHz}$ is the radio luminosity in W\,Hz$^{-1}$ derived from the 
MGCLS 1.28 GHz total flux density, using a power-law scaling and assuming a non-thermal 
spectral index of -0.8 \citep{Condon92}. $L_{\rm c}=6.4\times10^{21} W\,$Hz$^{-1}$is taken to be the
typical radio luminosity of an
$L_{*}$-like galaxy. \citet{Bell2003} argues that galaxies with low luminosities 
could have their non-thermal emission significantly suppressed and therefore need 
to be separated from the population with higher luminosities.

\subsection{Star formation activity as a function of clustercentric distance}
\label{sec:fsf_r200}

Previous studies have shown that SF 
activity is suppressed with decreasing distance from the
cluster centre \citep{Balogh1998,Lewis_2002,Haines_2015}.
We study the population distribution of
our radio-derived SFR for the 20 clusters with respect
to their clustercentric radius in units of $R_{200}$. We define the fraction
of star-forming galaxies as $f_{\rm SF}=N_{\rm SF}/N_{\rm tot}$, where
$N_{\rm SF}$ is the number of galaxies classified as star-forming cluster
members and $N_{\rm tot}$ is the total number of cluster members.
The left panel of Figure \ref{fig:frac_haines} shows the $f_{\rm SF}$
for MeerKAT-detected sources in radial bins
compared with $f_{\rm SF}$ results from \citet{Haines_2015}, obtained from 
infrared-derived SFR using \textit{Spitzer}/MIPS 24$\mu$m 
observations for 30 clusters at $0.15<z<0.30$ and a 
$5\sigma$ detection limit of $2\, \text{M}_{\odot}\text{yr}^{-1}$. The \citet{Haines_2015}
study calculates SFR using the \citet{Kroupa_2002} IMF, which yields results
that are nearly identical to the \citet{Chabrier_2003} IMF.
The fraction of star-forming galaxies from radio observations is generally higher than the $f_{\rm SF}$ of \citet{Haines_2015} with the data points having differences ranging between 
1-3$\sigma$. The observed higher $f_{\rm SF}$ in our sample could be a result of a bias towards merger-linked clusters. The \citet{Haines_2015} cluster sample is comprised
of uniformly selected clusters with no significant bias towards either merger or relaxed
clusters. They notice that even at large radii, 2$R_{200}$, the 
$f_{\rm SF}$ of clusters ($\approx0.23$) remained well below that of
field galaxies ($f_{\rm SF}=0.33\pm0.01$).
They conclude that this is not due to cluster galaxies having higher stellar masses 
than field galaxies, 
as it is known for $f_{\rm SF}$ to decrease with increasing stellar mass
\citep[e.g.,][]{Haines_2007}. They suggest that $f_{\rm SF}$ being low
out to large clustercentric radii may be due to
the star-forming galaxies being pre-processed 
in lower-density environments such as galaxy groups before they enter cluster environments.

The central panel of Figure \ref{fig:frac_haines} shows the $f_{\rm SF}$ of the MeerKAT 
cluster sample split by whether they have radio haloes/relics. 
Our sample has 14 clusters with extended emission in the form of radio haloes and
relics (found in merger clusters) and their $f_{\rm SF}$ trend is 
shown in blue stars. The $f_{\rm SF}$ trend for the 6 relaxed clusters
without radio haloes/relics is shown in 
green circles. The relaxed clusters have a shallower decline from
$1.9R_{200}$ to the centre in contrast to the steeper decline noticed
in the merging clusters. Recent studies \citep[][]{Cohen_2014,Cohen_2015,Yoon_2020} 
estimate that the fraction of star-forming galaxies is
$20-30\%$ higher in merger clusters than in relaxed clusters. The 
median $f_{\rm SF}$ for clusters with merger activity ($0.148\pm0.016$) in our sample is $\approx23\%$
higher than the relaxed clusters ($0.120\pm0.011$) for all star-forming galaxies within 
$R_{\rm 200}$. The difference rises to $\approx34\%$ for all star-forming galaxies
within 2$R_{200}$ between merger clusters ($0.225\pm0.019$) and relaxed clusters
($0.168\pm0.012$). The $1\sigma$ errorbars are calculated using bootstrap
resampling of the SF galaxies in the clusters.

\defcitealias{Popesso_2012}{Po12}
\defcitealias{Haines_2013}{Ha13}

\begin{figure}
\includegraphics[width=\columnwidth]{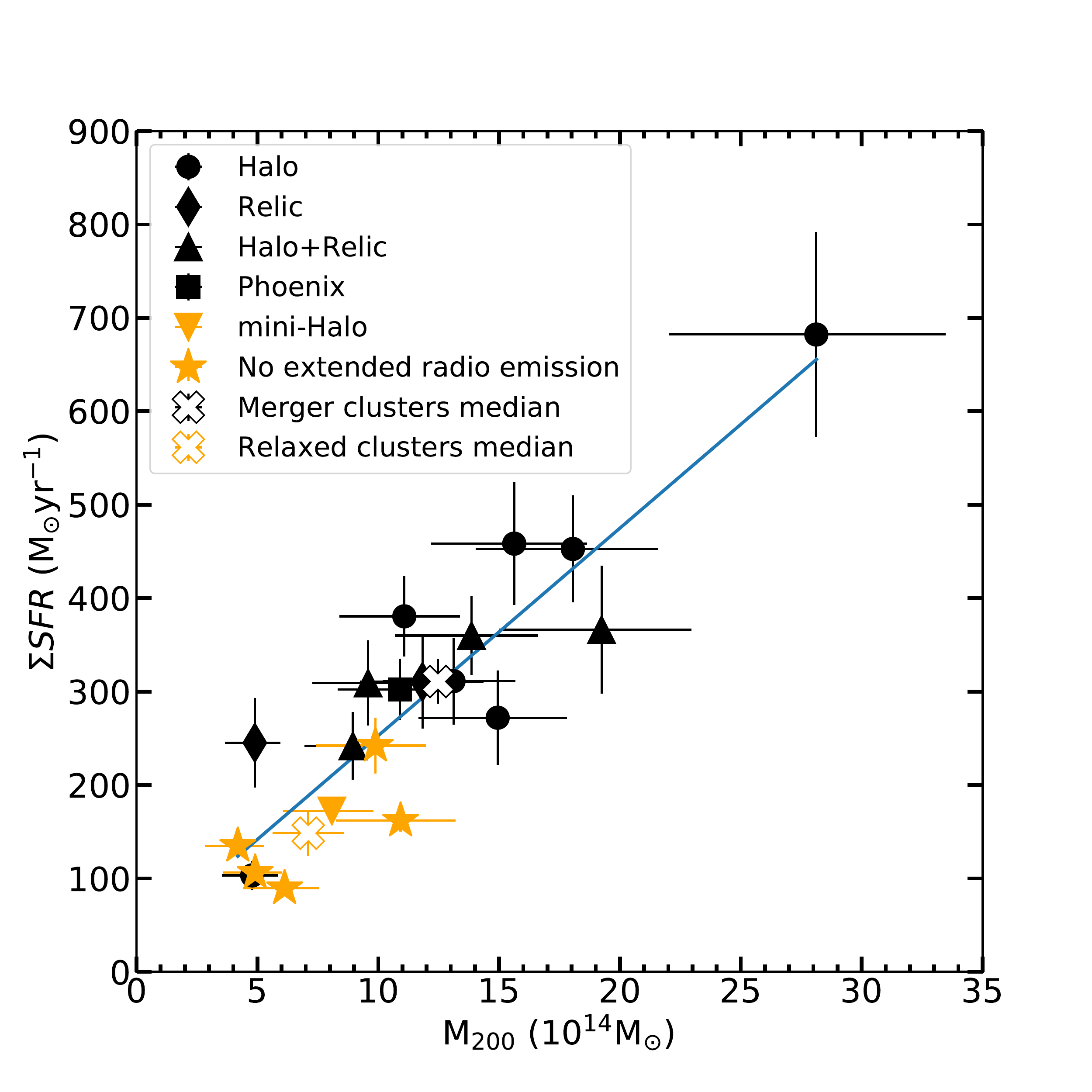}
\caption{Total  SFR in MGCLS clusters 
plotted against their cluster masses.
Clusters with extended radio emission linked to merger activity are
plotted in black markers (Clusters hosting only haloes are plotted in dots, 
those hosting haloes and relics are in upward-facing triangles and the ones with
only relics and a phoenix are in diamonds and squares respectively). Relaxed clusters
are plotted in orange markers (relaxed clusters with no diffuse emission are shown
in star markers and the mini-halo is indicated by a downward-facing triangle).
The error bars indicate $1\sigma$ uncertainties. The median values for
relaxed and merger clusters are shown by the X-shaped markers with
bootstrapped resampled errorbars. The blue line shows the least-squares line of best fit derived from the full cluster sample between the mass limits $4.19<M_{200}(10^{14}M_{\odot})<28.13$}
\label{fig:totSFRvsMass}
\end{figure}

\begin{figure}
\includegraphics[width=\columnwidth]{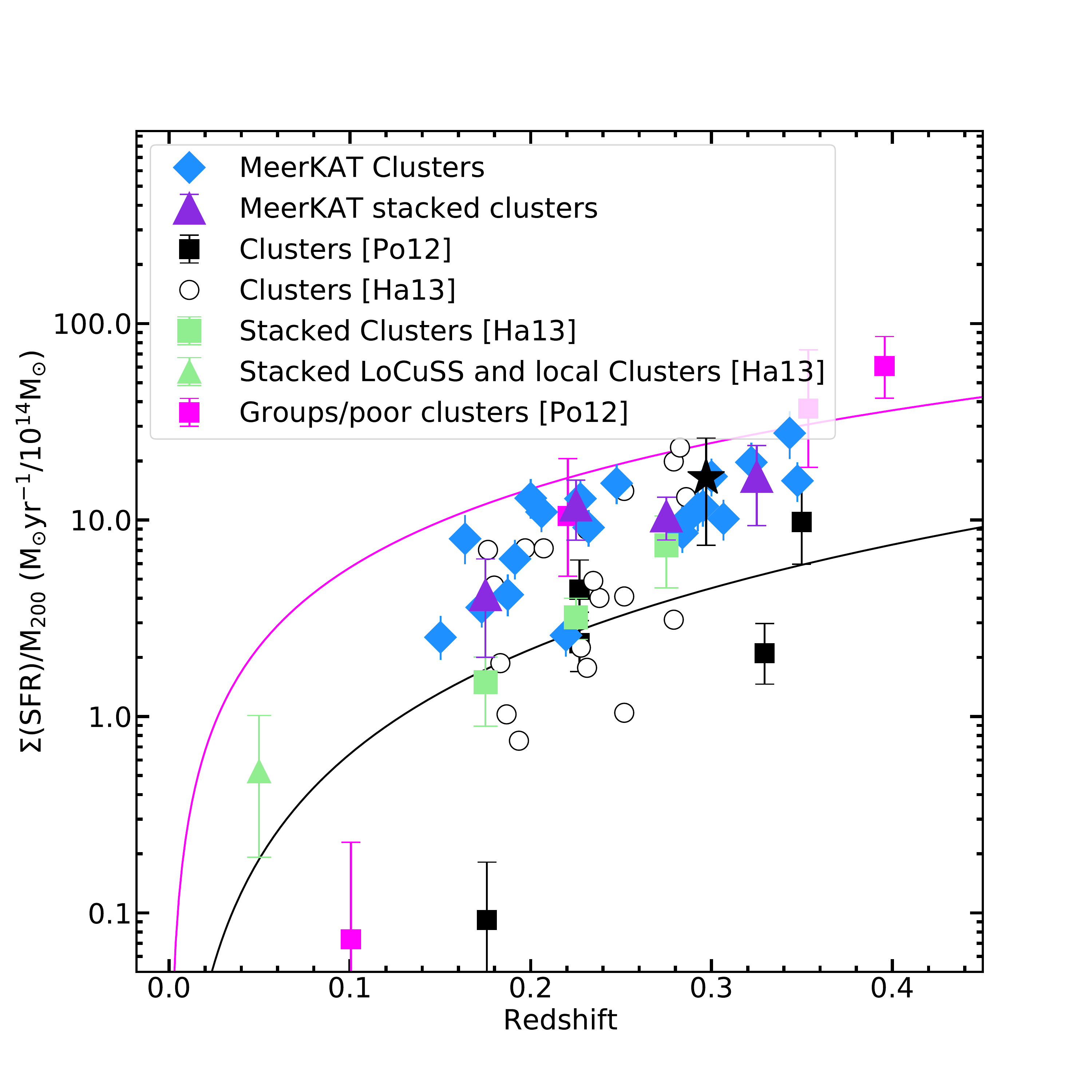}
\caption{The Evolution of $\Sigma(\text{SFR})/M_{200}$ - 
redshift relation among cluster LIRGs.
The blue diamonds indicate the total SFR normalised by the cluster mass
for all the cluster LIRGs within $R_{200}$ for each
MGCLS cluster as a function of redshift. The
purple triangles indicate the averaged values for 
the MGCLS clusters in four redshift bins in steps
of 0.5. There are 4 clusters at $0.15<z<0.20$,
6 clusters at $0.20<z<0.25$, 4 clusters at $0.25<z<0.30$
and 5 clusters at $0.30<z<0.35$. We use bootstrap 
resampling in the $\Sigma(\text{SFR})/M_{200}$ scatter 
of clusters in each bin to estimate $1\sigma$
uncertainties. The \citetalias{Haines_2013}
clusters are indicated by black rings and
their redshift binned average values are shown
in green squares. \citetalias{Haines_2013} also included the analysis of local clusters at $0.02<z<0.05$ shown by the green triangle. The black and magenta squares are from the 
\citetalias{Popesso_2012} clusters and groups 
respectively which are made up of composite systems from
GOODS, COSMOS and individual systems. The black
star indicates the merging Bullet cluster
also from \citetalias{Popesso_2012}. The black 
curve shows the best fit $\Sigma(\text{SFR})/M_{200}$
relation for the \citetalias{Popesso_2012} cluster
sample excluding the Bullet cluster and the 
magenta curve shows the relation for the group/poor clusters sample.
The \citetalias{Popesso_2012} clusters and groups best-fit 
lines were deduced from their study
of clusters and groups at $0.1<z<1.6$. For our study, we plot the 
\citetalias{Popesso_2012} best-fit lines at the part 
that highlights our redshift slice.}
\label{fig:sfrM}
\end{figure}

To investigate if our results could have a 
hidden dependence on cluster mass since all relaxed clusters in our sample
have masses below the median mass, we divided our sample into two mass bins 
split by median mass ($1.09\times10^{15}\,M_{\odot}$). The right panel
of Figure \ref{fig:frac_haines} shows the $f_{\rm SF}$ trends for 10 clusters below the median mass
split by their dynamical state, with 5 relaxed clusters (green circles) and 5 clusters with radio 
haloes/relics (brown diamonds). The remaining 10 clusters above the median mass host radio
haloes/relics are shown in blue stars. The $f_{\rm SF}$ trend of merger-linked clusters
below the median mass shows more SF activity than seen in the trend of the relaxed clusters in the 
same mass bin, roughly consistent with the $f_{\rm SF}$ trend of merger-linked clusters in the 
higher mass bin. This enhanced SF activity in merger-linked clusters below median mass is also evident in 
Figure \ref{fig:totSFRvsMass} which is discussed in the following section.

We notice similar $f_{\rm SF}$ trends within $1\sigma$ between clusters 
that host only haloes and the clusters that host only relics and 
those with both relics and haloes in our sample. 

\subsection{SFR -- \texorpdfstring{{$M_{200}$\,\,}} rrelations}

\subsubsection{Total star formation rates versus cluster mass}
\label{sec:totSFR_M}
To investigate how our cluster sample SFR relates with cluster mass,
we summed the SFR for member galaxies within $R_{200}$ and 
plotted them against cluster masses as shown in Figure \ref{fig:totSFRvsMass}.
The clusters that have been identified to have merger-related extended
radio emission are plotted in black markers and relaxed clusters are 
plotted in orange markers.
Figure \ref{fig:totSFRvsMass} shows that the total SFR correlates not 
only with cluster mass but is also dependent upon the dynamical state of the cluster. 
This result is consistent with recent claims that mergers enhance SF activity in
clusters at low redshifts \citep[e.g.,][]{Cassano_2013,Cohen_2014,Stroe_2021}.
We observe a marginal difference (within $3\sigma$) between median values for mass normalised total
SFR in relaxed clusters ($21.5\pm1.9$\,M$_{\odot}$yr$^{-1}$/$10^{14}$M$_{\odot}$)
and clusters with merger activity ($26.1\pm1.6$\,M$_{\odot}$yr$^{-1}$/$10^{14}$M$_{\odot}$).
The errors on the median values were estimated from bootstrap resampling.

\subsubsection{Star formation activity per unit cluster mass}
\label{sec:sfr_M}

A number of recent studies have examined the evolution
of SF activity with redshift in clusters by using the total SFR
per cluster mass for clusters hosting luminous infrared galaxies (LIRGs) 
\citep[e.g.,][]{Popesso_2012,Haines_2013,Webb_2013}.
This is done by summing the SFR of 
individual galaxies that are confirmed to be members within $R_{200}$
for each cluster and normalising it by $M_{200}$, $\Sigma(\text{SFR})/M_{200}$. 
This approach enables the comparison of SF activity for clusters with different masses. 
Although methods employed by the various studies to define cluster 
properties differ from one another, including this one, they all consistently come to the
conclusion of a rapid decline in SF activity within clusters with decreasing redshift.

To understand the evolution of SF activity within our sample, we follow the analysis
of \citet[][]{Popesso_2012}, hereafter \citetalias{Popesso_2012}, which studies the $\Sigma(\text{SFR})/M_{200}$ -- redshift relation
for LIRGs in rich/massive clusters
and groups/poor clusters at $0.1<z<1.6$ using infrared luminosities from
\textit{Herschel}/PACS 100$\mu$m and 160$\mu$m. 
Following \citetalias{Popesso_2012}, we summed up the  SFR 
of all the MeerKAT galaxies with SFR values above the LIRG 
luminosity limit corresponding to 
$L_{\rm 1.4GHz}\approx3.14\times10^{22}\,W$\,Hz$^{-1}$ or 10\,M$_{\odot}$yr$^{-1}$ 
in SFR for
each of the 20 clusters in our sample. All clusters in our sample
contain at least one LIRG within $R_{200}$. 
Figure \ref{fig:sfrM} shows a plot of $\Sigma(\text{SFR})/M_{200}$ -- redshift relation
for our sample including results from \citetalias{Popesso_2012} scaled down by
1.74 from the Salpeter IMF to Chabrier IMF. Figure \ref{fig:sfrM} includes results
from \citet[][]{Haines_2013}, hereafter \citetalias{Haines_2013}, which also 
followed the \citetalias{Popesso_2012} study of LIRGs 
at $0.15<z<0.3$ using infrared luminosities from \textit{Spitzer}/MIPS 24$\mu$m.
It is worth noting that there are differences in cluster mass
estimation methods between our study and both the \citetalias{Popesso_2012} 
and the \citetalias{Haines_2013} studies. The \citetalias{Popesso_2012} study
uses dynamical mass estimates derived from optical spectroscopy and the
\citetalias{Haines_2013} study relies on X-ray-derived masses. This means that
for clusters that are common among the studies, the total SFR 
within $R_{\rm 200}$ would be estimated out to different radii, due to
differences in the mass measurements between the studies.
None of the clusters in our sample overlap with the \citetalias{Popesso_2012}
cluster sample and only one cluster from our sample (Abell 209) appears in
the \citetalias{Haines_2013} study. The $\Sigma(\text{SFR})/M_{200}$ value
for Abell 209 from \citetalias{Haines_2013} is $\approx7.1$  
M$_{\odot}$yr$^{-1}$/$10^{14}$M$_{\odot}$.
This is within $1.5\sigma$ of
our estimated value for the same cluster, ($10.99\pm2.69$) M$_{\odot}$yr$^{-1}$/$10^{14}$M$_{\odot}$.

We estimate the overall evolutionary trend by dividing 
our cluster sample into four redshift bins
as indicated by the purple triangles in Figure \ref{fig:sfrM}. 
We observe a $\approx4\times$ decline in the
level of SF activity per cluster mass in MeerKAT 
clusters in contrast to the $\approx5\times$
decline seen by \citetalias{Haines_2013} 
in green squares over the $0.15<z<0.3$ redshift slice. 
Just as observed in \citetalias{Haines_2013},
we also see a decline in the number of cluster LIRGs with 
redshift bins, reducing from 87 LIRGs in 5 cluster systems
at $0.3<z<0.35$ to 10 LIRGs in 5 systems at $0.15<z<0.20$.

Figure \ref{fig:sfrM} includes the $\Sigma$(SFR)/$M_{\rm 200}$ of the
Bullet cluster from \citetalias{Popesso_2012} (indicated by the black star), 
which is known to have an ongoing violent merger. The Bullet cluster has
a higher $\Sigma$(SFR)/$M_{\rm 200}$, consistent within errorbars with the clusters
in our sample, most of which host extended radio emission linked to merger
activity. Even with noted differences in our study and those from other
authors, the evolutionary trend for MGCLS clusters also 
indicate that there is a rapid decline in star 
formation activity among cluster galaxies at lower redshifts, roughly consistent with 
observations made by \citetalias{Popesso_2012} and \citetalias{Haines_2013}.

\section{Discussion}
\label{sec:discussion}

In the previous section, we studied the effect of cluster dynamical
state on the SF activity of galaxies in relaxed clusters and clusters with 
merger-linked extended emission. Our results infer that clusters with
merger-linked extended emission have enhanced SF activity compared to relaxed clusters.
We estimate a $23\%$ higher SF activity in clusters with radio haloes and/or relics
within $R_{\rm 200}$ and $34\%$ out to 2$R_{\rm 200}$. We note that our dynamical state subsamples are unevenly distributed across the full sample redshift range as shown in Figure \ref{fig:z_bins}. The relaxed clusters fall within the lower redshift bin while unrelaxed/merger clusters are all above $z=0.2$. The $f_{\rm SF}$ in clusters has been shown to increase with redshift, forming a relation known as the \citet[][hereafter BO]{ButcherOemler_1978} effect. The BO effect is the increase in the fraction of blue galaxies in clusters with redshift. Since our study uses $f_{\rm SF}$ to compare SF activity between subsamples, we expect some fraction of the differences in the overall cluster $f_{\rm SF}$ to be due to the BO effect. However, as suggested by this investigation as well as previous studies that have investigated SF activity in clusters, the dynamical state of a cluster plays an influential role in its SF activity. Observing a homogeneous sample (not subject to MGCLS heterogeneous nature) will allow us to confirm this result and rule out whether or not the BO effect affected our results and by how much.

\begin{figure}
\includegraphics[width=\columnwidth]{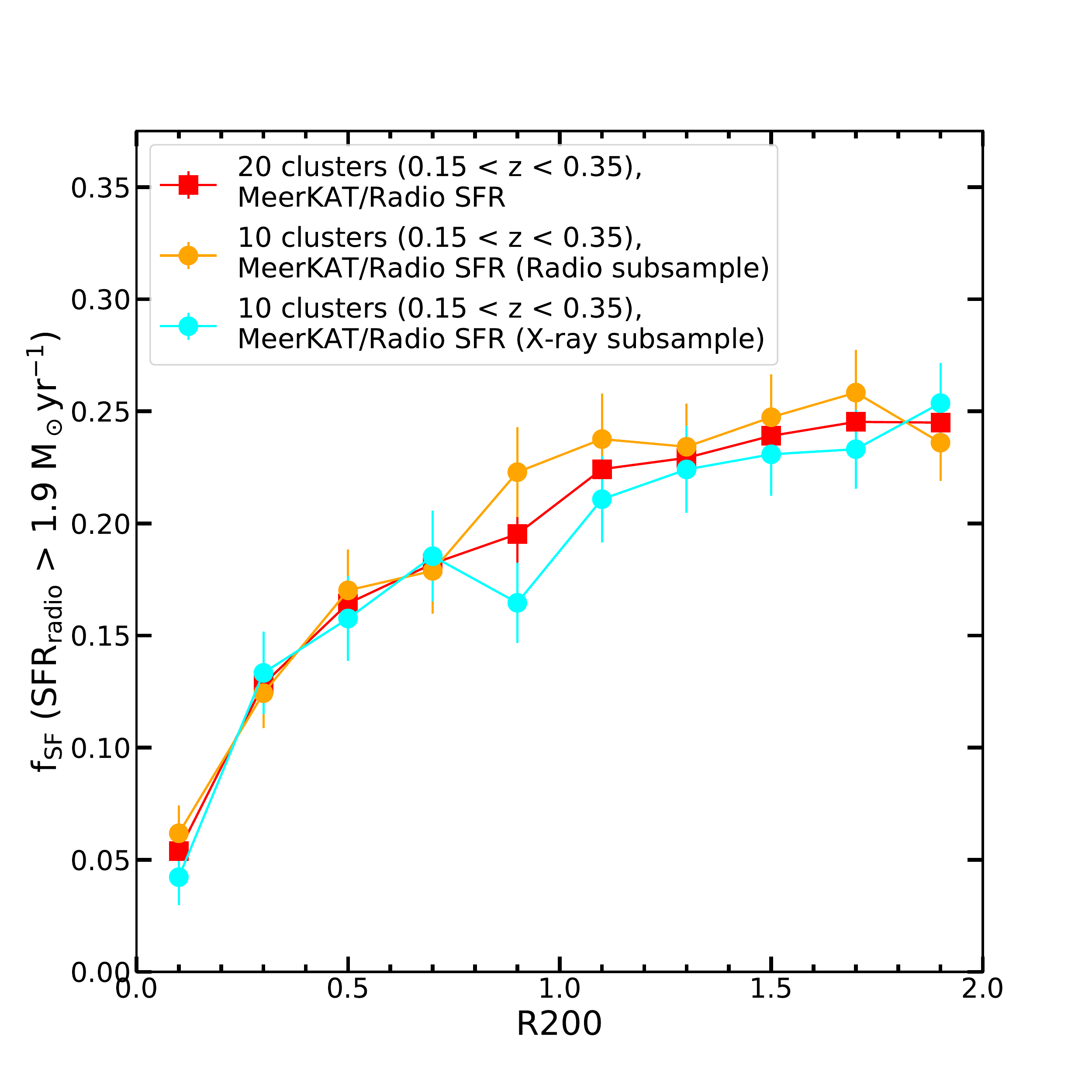}
\caption{The fraction of star-forming galaxies including trends for clusters split by their
MGCLS subsamples. The $f_{\rm SF}$ trend for clusters from the radio-selected subsample is plotted
in orange dots and the trend for clusters from the X-ray-selected subsample is plotted
in cyan dots.}
\label{fig:fsf_subsample}
\end{figure}

\begin{figure*}
    \includegraphics[width=0.495\textwidth]{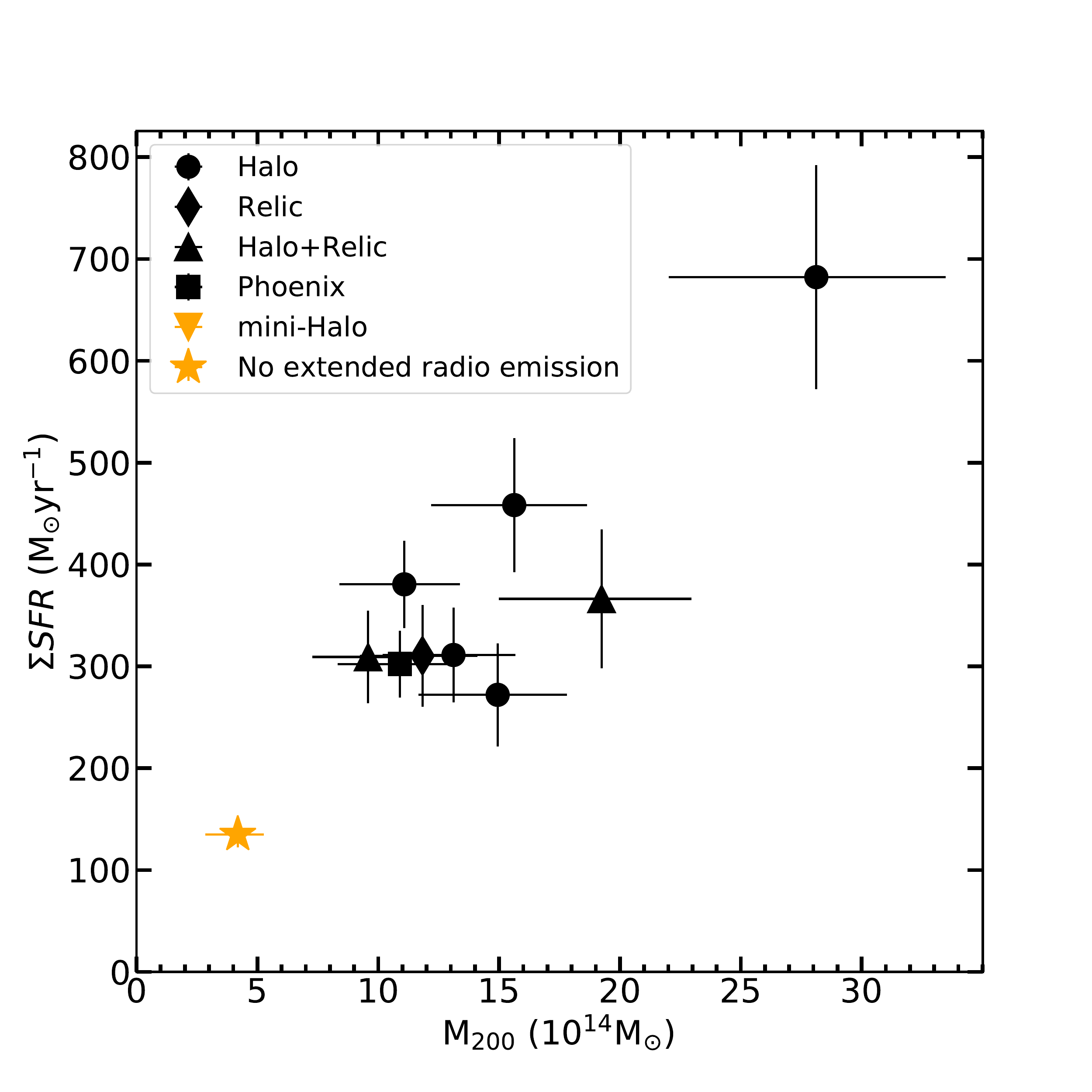} 
    \includegraphics[width=0.495\textwidth]{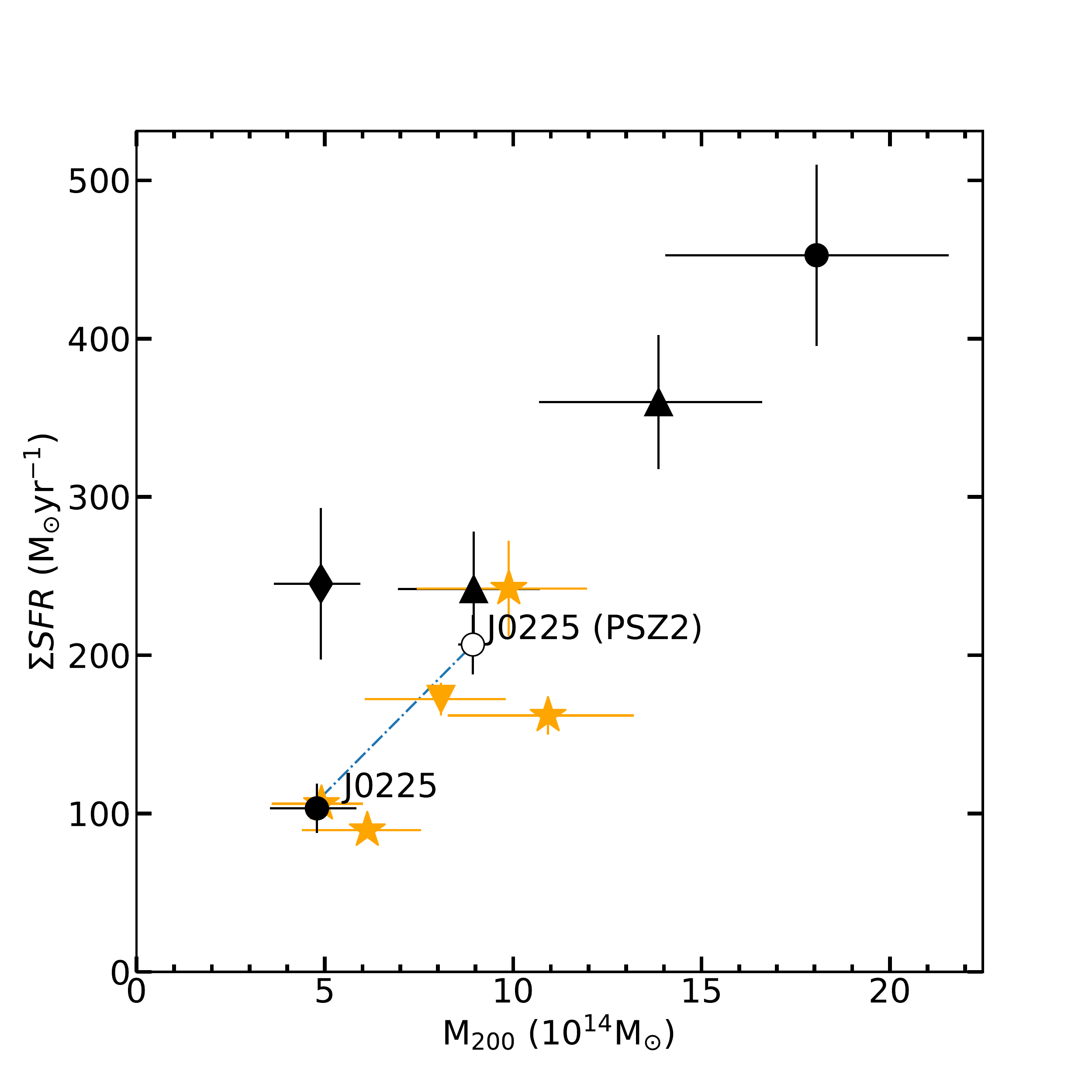} 
        
\caption{$\Sigma{\rm SFR}$ -- $M_{200}$ relation for 
clusters from the MGCLS radio-selected (\textit{left})
and the X-ray-selected (\textit{right}) subsamples. The open symbol shows 
the $\Sigma{\rm SFR}$ -- $M_{200}$
for cluster J0225.9-4154 (J0225) at the Planck SZ mass estimate from the
Planck Sunyaev-Zeldovich sources catalogue \citep[PSZ2,][]{Planck_2016}. 
The dashed blue line shows the difference between the $\Sigma{\rm SFR}$ -- $M_{200}$
from the ACT mass estimate and the PSZ2 mass estimate.}
\label{fig:totSFRvsMass_sub}
\end{figure*}

Results from our study are consistent with observations made by previous studies that looked at
the effect of cluster dynamical state on SF activity.
\citet{Cohen_2014} studied 107 clusters at $0.04<z<0.1$ using the Sloan Digital Sky Survey (SDSS)
and uses substructure within clusters to identify clusters with
multiple substructure components (mergers) from single component clusters (relaxed).
They found merger clusters to have $\approx30\%$ more SF activity than relaxed clusters.
\citet{Yoon_2020} estimates up to $\approx24\%$ more SF activity within $R_{\rm 200}$ in 
multiple component clusters compared to single component clusters
from their study of 105 clusters at $0.015<z<0.060$ using SDSS.
The enhanced SF activity in merger clusters has also been observed
by \citet{Stroe_2021} from their study of H$\alpha$ emitters in 14 clusters at $0.15<z<0.31$.

In contrast to the results of these low-redshift studies, \citet{Mansheim_2017}
observed that SF activity may be suppressed in merger clusters
from the study of an individual cluster system at high redshift ($z\approx1.105$). More recently, \citet{Maier_2022} observed higher SF activity in clusters with an actively star-forming BCG (indicative of a relaxed cool-core cluster) than in clusters with passive BCGs (indicative of a non-relaxed cool-core cluster) using a study of 18 clusters at $0.15<z<0.26$ from the Local Cluster Substructure Survey (LoCuSS).
The \citet{Chung_2010} study of the merging Bullet cluster argues that, depending on the cluster,
mergers may not be the driver for the enhanced SF activity but could be a result of the infalling
galaxy population. \citet{Wittmann_2019} suggests that claims made by various studies that
cluster mergers suppress, enhance or have no effect on SF activity may not be conflicting
but can be explained by how much time has passed since the pericenter passage (the age of the merger), the relative velocity at the first pericenter, and the viewing angle.

We observe a linear correlation between $\Sigma{\rm SFR}$ and cluster mass ($M_{200}$),
consistent with results from previous studies \citep[e.g.,][]{Goto_2004,Popesso_2006}.
This correlation has been described as a richness effect, $\Sigma{\rm SFR}\propto{N_{\rm gal}}$,
where $ N_{\rm gal}$ is the number of galaxies in a cluster. The number of galaxies in a cluster
scales with cluster mass in that the higher the mass, the higher the number of star-forming galaxies.
Figure \ref{fig:totSFRvsMass} shows the linear correlation of $\Sigma{\rm SFR}$ and $M_{200}$.
We also observe the contribution of cluster dynamical state to the $\Sigma{\rm SFR}$, something that
was not considered by previous studies.
The contribution of cluster dynamical state is particularly evident
along clusters with mass estimates within $1\sigma$ of each other but with different
relaxation states. This difference is noticeable in clusters within the mass range
$0.8-1.2\times10^{15}\, M_{\odot}$. There are 8 clusters that fall within this range and
we observe that the median $\Sigma{\rm SFR}$ for relaxed clusters ($172.3\pm12.1$\,M$_{\odot}$yr$^{-1}$) is $\approx0.5\times$ lower than the one for clusters with radio haloes and/or relics ($309.2\pm42.9$\,M$_{\odot}$yr$^{-1}$).

As described in Section \ref{sec:sample}, the MGCLS catalogue is made up of a combination of two
subsamples. One from the radio-selected subsample
(biased towards extended emission) and the other from the X-ray-selected subsample
with no prior biases towards or against clusters with extended emission but selected
with no redshift or X-ray luminosity criteria followed. Our cluster
sample is biased towards clusters with extended radio emission, albeit equally split between 
clusters from each MGCLS subsample. To check our results, we look at clusters from
the X-ray-selected subsample by splitting clusters in our sample by their MGCLS 
subsamples and re-plotting Figures \ref{fig:frac_haines} and \ref{fig:totSFRvsMass}.

Figure \ref{fig:fsf_subsample}
shows the $f_{\rm SF}$ profiles of our clusters broken down by their subsample. 
The radio-selected subsample comprises 9 merger-linked extended emission clusters and one relaxed
cluster in the redshift range $0.15<z<0.35$. The X-ray-selected clusters subsample 
is evenly distributed between 5 merger-linked extended emission 
clusters and 5 relaxed clusters also in the redshift range of $0.15<z<0.35$. The 
radio-selected subsample has an $f_{\rm SF}$ of $0.253\pm0.019$ 
within 2$R_{\rm 200}$ while the
X-ray-selected subsample is $0.241\pm0.012$ out to the same radius.
The $1\sigma$ errorbars are calculated using bootstrap
resampling of the SF galaxies in the clusters. The radio-selected $f_{\rm SF}$ 
trend is slightly higher than that of the X-ray-selected sample and the $f_{\rm SF}$ values 
lie within $1-1.5\sigma$ between the subsamples. The observed higher $f_{\rm SF}$ 
trend of the radio-selected subsample is likely due to the selection bias towards clusters with haloes
and/or relics, and therefore to massive mergers.

Figure \ref{fig:totSFRvsMass_sub} shows the plot of $\Sigma{\rm SFR}$ versus $M_{200}$
for clusters in our sample separated into each respective subsample. The left panel shows the plot
of the clusters from the radio-selected subsample and the right panel shows the plot of the
clusters from the X-ray-selected subsample. Although clusters from the X-ray-selected 
sample is limited in quantity, a correlation between $\Sigma{\rm SFR}$ and $M_{200}$ is observed
in the right panel of Figure \ref{fig:totSFRvsMass_sub}. Results from the X-ray-selected
clusters are consistent with merger clusters having higher SF activity than relaxed clusters.
Due to the limitation of a small cluster sample, future observations of a larger sample
will be required to confirm these results.

We observe one halo-hosting cluster,
J0225.9-4154 (J0225 hereafter),
that has a $\Sigma{\rm SFR}$ profile of relaxed clusters at its mass on the X-ray subsample.
Dynamical state studies of J0225 reveal that it is an interesting binary cluster
system of two merging clusters (A3017 and A3016) connected by an X-ray filament 
\citep[see][for detailed reviews of the on-going merger]{Foex_2017,Parekh_2017,Chon_2019}.
Dynamical studies show that J0225 is a face-on merger at its early stage with multiple
substructures within its virial radius.
The dynamical mass estimate ($M_{200}$) for J0225 according to \citet{Foex_2017} is
$18.6\substack{+2.6\\-2.0}\times10^{14}\,M_{\odot}$
but after accounting for the substructure in the line of sight they obtain an estimate
of $8.4\substack{+1.3\\-1.5}\times10^{14}\,M_{\odot}$. Their estimate is in good agreement with the
mass estimate from the Planck Sunyaev-Zeldovich sources catalogue \citep[PSZ2,][]{Planck_2016}
re-scaled to $M_{200}$, $8.9\pm0.4\times10^{14}\,M_{\odot}$.
The \citet{Foex_2017} and the PSZ2 mass estimate for J0225 are approximately double the 
ACT mass estimate ($4.8\substack{+1.2\\-1.0}\times10^{14}\,M_{\odot}$) but fall 
within $3\sigma$ of each other. The right panel of Figure 
\ref{fig:totSFRvsMass_sub} includes the data point for
$\Sigma{\rm SFR}$ at the PSZ2 mass estimate for J0225 and shows the difference
in $\Sigma{\rm SFR}$ between the two mass estimates of the cluster. The
$\Sigma{\rm SFR}$ from the PSZ2 mass estimate is higher by a factor of 1.6 and has 
$R_{200}$ at 1.9 Mpc, $1.3\times$ higher than $R_{200}$ from the ACT mass estimate.
The extraordinary nature of J0225 raises the need for a multi-wavelength 
analysis to get a clear picture of its properties.
\section{Conclusions}
\label{sec:conclusion}

We have studied SF activity 
in cluster environments at low redshifts
($0.15<z<0.35$) using dust-unbiased 
radio continuum data from the MGCLS.
This provides the first-look at 
SFR in clusters using
radio data from the MeerKAT telescope. 
The investigation of SF activity in and around clusters is crucial
in understanding how galaxies evolve with time in high-density environments and how
the dynamical state of a cluster affects SF activity in galaxies.
Below is the summary of our main results:

\begin{enumerate}
\item We measure a $f_{\rm SF}$ population trend out to a similar limit at $2R_{200}$ with the \citet{Haines_2015} study, noting a 1-3$\sigma$ difference between the radio SFR and IR SFR data points which is likely due to the MGCLS selection bias towards merger-linked clusters. Both the radio-derived and IR-derived $f_{\rm SF}$ are 
lower than the \citet{Haines_2015} field galaxies $f_{\rm SF}$ even at 2$R_{200}$. 
This is consistent with the suggestion that in-falling galaxies may be pre-processed 
in their prior environments such as galaxy groups, 
before they enter cluster environments, which could explain
the lower $f_{\rm SF}$ even at radii over $2R_{200}$.

\item There is a difference in SF activity between clusters that host 
extended radio emission (relics and haloes) linked to cluster mergers 
and clusters that are non-halo/relic hosting, likely to be dynamically relaxed
or only minor mergers. 
We see the differences in the $f_{\rm SF}$ as well as the $\Sigma{\rm SFR}$ 
between merger clusters and relaxed clusters. Merger clusters have 
a higher fraction of star-forming galaxies and consequently a higher 
$\Sigma{\rm SFR}$ compared to relaxed clusters. 

\item We find a rapid decline in the SF evolutionary trend among radio-selected
galaxies with SFR corresponding to those of cluster
LIRGs in our sample. We observe a $\approx4\times$ decline in $\Sigma{\rm SFR}/M_{200}$
from redshift of 0.35 till 0.15, corresponding to 2\,Gyr in lookback time.
This observation is roughly consistent with IR-derived SFR studies
by \citet{Popesso_2012} and \citet{Haines_2013}.
    
\end{enumerate}

One limitation of this work is that the underlying sample (MGCLS) is heterogeneous 
and made up of two subsamples, with one biased towards clusters with extended emission
and the other probing  X-ray clusters with no selection criteria followed.
A follow-up study will be conducted in the future
using a catalogue of unbiased cluster samples selected only by mass
from the MeerKAT Exploration of Relics, Giant Halos, and Extragalactic 
Radio Sources \citep[MERGHERS program, see][]{Kenda_2021}.

\section*{Acknowledgements}
We are grateful to the referee (Christopher P. Haines) for their invaluable comments and suggestions that improved the clarity of the paper. The authors would like to thank Lawrence Rudnick and Bruce Partridge for useful discussions that helped to improve the quality of this work. The MeerKAT telescope is operated by the South African Radio Astronomy Observatory, is a facility of the National Research Foundation, an agency of the Department of Science and Innovation. The authors acknowledge the contribution of those who designed and built the MeerKAT instrument. The National Radio Astronomy Observatory is a facility of the (US) National Science Foundation, operated under a cooperative agreement by Associated Universities, Inc.

KK acknowledges funding support from the South African
Radio Astronomy Observatory (SARAO) through the HCD scholarship.
Computations were done on the hippo cluster at the University of KwaZulu Natal.

\section*{Data Availability}

The MeerKAT 1.28 GHz radio continuum data is accessed from the MeerKAT Galaxy Clusters Legacy Surveys 
(MGCLS\footnote{\url{https://archive-gw-1.kat.ac.za/public/repository/10.48479/data_releases/The_MeerKAT_Galaxy_Cluster_Legacy_Survey_DR1/index.html}}) data archive. 
The data generated in this study will be shared on reasonable request to the corresponding author and will be made
public through the MGCLS data archive in the future.



\bibliographystyle{mnras}
\bibliography{MGCLS_SFRs} 




\appendix

\section{Legacy Surveys images}

Here we present Legacy Surveys DR9 images for our cluster sample and the 
identified radio-loud galaxies within $R_{\rm 200}$. See discussion
in Section \ref{sec:data}.

\begin{figure*}
\centering
\vspace{-0.5cm}
\includegraphics[trim=2cm 2cm 2cm 0.cm,scale=1.]{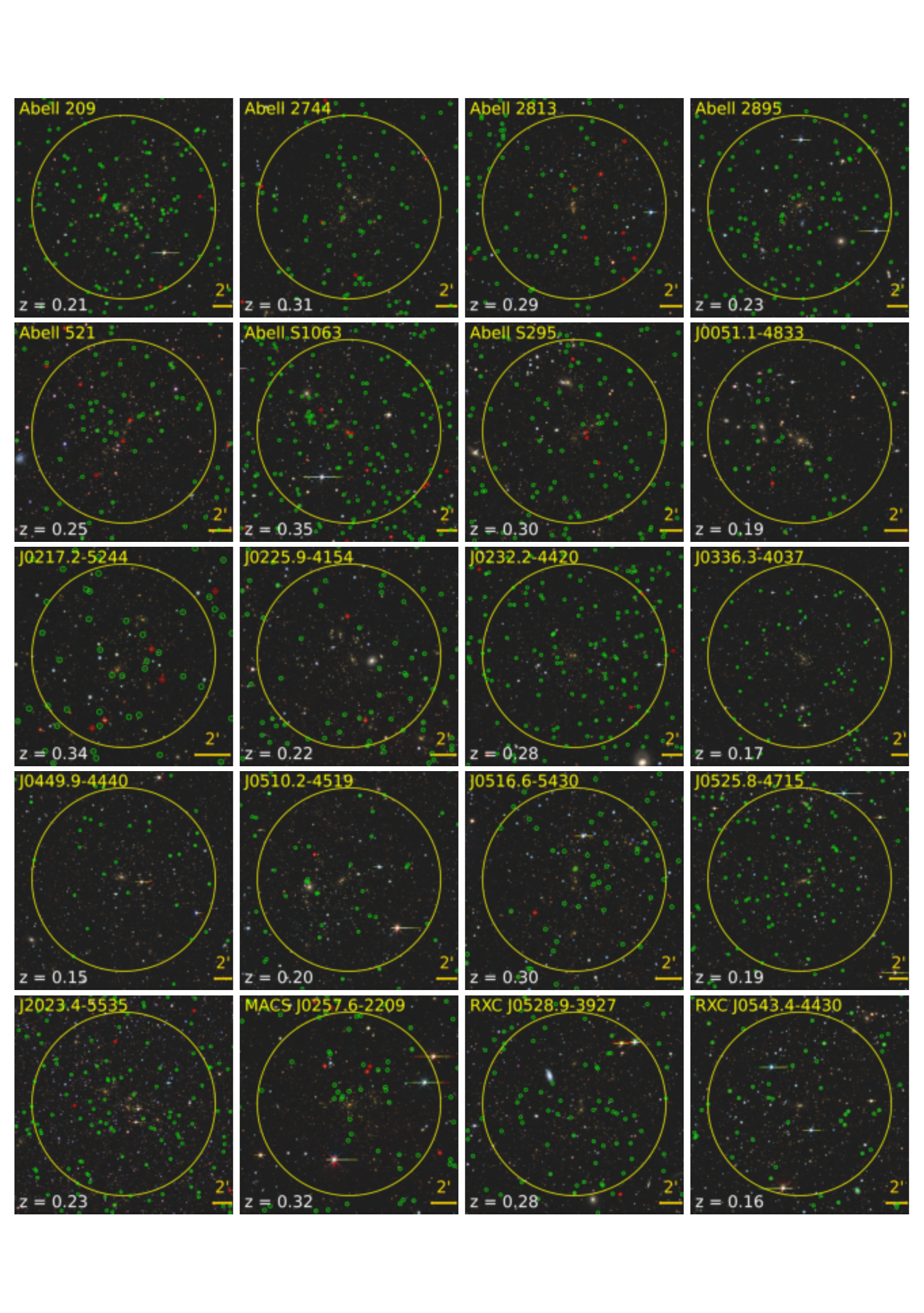}
\caption{Legacy Surveys DR9 images of the 20 clusters with radio detected
galaxies highlighted in small green and red circles representing star-forming galaxies 
and radio-loud AGN respectively. 
The large yellow circles show the $R_{200}$ of the cluster.}
\label{fig:multi_panel}
\end{figure*}

Figure \ref{fig:multi_panel} shows the Legacy Surveys DR9 images of our cluster sample with the identified MeerKAT-detected cluster members.

\bsp	
\label{lastpage}
\end{document}